\documentclass[12pt,fleqn,a4paper]{article} % A4 size
\usepackage{epsfig,graphics,graphicx}
\usepackage{clshan-math,clshan-dm-dd}
\def \lsim {\:\raisebox{-0.7ex}{$\stackrel{\textstyle<}{\sim}$}\:}
\def \gsim {\:\raisebox{-0.7ex}{$\stackrel{\textstyle>}{\sim}$}\:}
\begin{document}
\thispagestyle{empty}
\begin{flushright}
 March 2011
\end{flushright}
\begin{center}
{\large\bf
 Effects of Residue Background Events
 in Direct Dark Matter \\ \vspace{-0.04cm} Detection Experiments on
 the Estimation of the \\ \vspace{0.1cm} Spin--Independent WIMP--Nucleon Coupling} \\
\vspace{0.7cm}
 {\sc Chung-Lin Shan} \\
\vspace{0.5cm}
 {\it Department of Physics, National Cheng Kung University \\
      No.~1, University Road,
      Tainan City 70101, Taiwan, R.O.C.}                    \\~\\
 {\it Physics Division,
      National Center for Theoretical Sciences              \\
      No.~101, Sec.~2, Kuang-Fu Road,
      Hsinchu City 30013, Taiwan, R.O.C.}                   \\~\\
 {\it E-mail:} {\tt clshan@mail.ncku.edu.tw}                \\
\end{center}
\vspace{1cm}
\begin{abstract}
 In our work on the development of
 a model--independent data analysis method
 for estimating the spin--independent (SI) scalar coupling of
 Weakly Interacting Massive Particles (WIMPs)
 on nucleons
 by using measured recoil energies
 from direct Dark Matter detection experiments directly,
 it was assumed that
 the analyzed data sets are background--free,
 i.e., all events are WIMP signals.
 In this article,
 as a more realistic study,
 we take into account
 a fraction of possible residue background events,
 which pass all discrimination criteria and
 then mix with other real WIMP--induced events
 in our data sets.
 Our simulations show that,
 for the estimation of the SI WIMP--nucleon coupling,
 the maximal acceptable fraction of residue background events
 in the analyzed data set
 of ${\cal O}(50)$ total events
 is \mbox{$\sim$ 10\% -- 20\%}.
 For a WIMP mass of 100 GeV
 and 20\% residue background events,
 the systematic deviation of the reconstructed SI WIMP coupling
 (with a reconstructed WIMP mass)
 would in principle be $\sim +13\%$
 with a statistical uncertainty
 of {$\sim^{+21\%}_{-14\%}$}
 ($\sim -3.3\%^{+18\%}_{-13\%}$
  for background--free data sets).
\end{abstract}
\clearpage
\section{Introduction}
 Currently,
 direct Dark Matter detection experiments
 searching for Weakly Interacting Massive Particles (WIMPs)
 are one of the promising methods
 for understanding the nature of Dark Matter (DM)
 and identifying them among new particles produced at colliders
 as well as reconstructing the (sub)structure of our Galactic halo
 \cite{Smith90, Lewin96, SUSYDM96, Bertone05}.
 To this aim,
 model--independent methods for
 determining the WIMP mass
 \cite{DMDDmchi-SUSY07, DMDDmchi}
 as well as
 estimating the spin--independent (SI) WIMP coupling on nucleons
 \cite{DMDDfp2-IDM2008, DMDDfp2}
 from direct detection experiments
 have been developed%
\footnote{
 In the literature,
 another method based on the maximum likelihood analysis
 has also been discussed
 \cite{Green-mchi, Bernal08, DMDDmchi-NJP}.
 However,
 in contrast to the model--independent procedures,
 this maximum likelihood analysis
 requires prior knowledge/assumptions
 about the velocity distribution function
 and the local density of halo WIMPs.
 The WIMP mass and the SI cross section on nucleons
 determined by this method
 are also coupled.
}.

 These methods built basically on the work
 on the reconstruction of the (moments of the)
 one--dimensional velocity distribution function of halo WIMPs
 by using experimental data
 (measured recoil energies) directly
 \cite{DMDDf1v}.
 The spectrum of recoil energy
 is proportional to an integral over
 the one--dimensional WIMP velocity distribution,
 $f_1(v)$,
 where $v$ is the absolute value of the WIMP velocity
 in the laboratory frame.
 In fact,
 this integral is just the minus--first moment of
 the velocity distribution function,
 which can be estimated from experimental data directly
 \cite{DMDDf1v, DMDDmchi}.
 Then,
 by assuming that
 the SI WIMP--nucleus interaction dominates
 and the WIMP couplings on protons and on neutrons
 are approximately equal,
 this SI WIMP coupling on nucleons
 can be estimated from experimental data directly
 \cite{DMDDfp2-IDM2008, DMDDfp2}.
 It was found that,
 by combining experimental data sets
 with different target nuclei,
 the SI WIMP--nucleon coupling can be estimated
 {\em without} making any assumption
 about the velocity distribution of halo WIMPs
 {\em nor} prior knowledge about the WIMP mass
 \cite{DMDDfp2-IDM2008, DMDDfp2}%
\footnote{
 Note that,
 as will be discussed in more details later,
 the WIMP mass and the local WIMP density
 are needed for this estimation.
 While the former can be determined
 from (other) direct detection experiments directly
 \cite{DMDDmchi-SUSY07, DMDDmchi},
 the latter has conventionally been estimated
 by means of the measurement of
 the rotation curve of the Milky Way
 with an uncertainty of a factor of $\sim$ 2
 \cite{SUSYDM96, Bertone05}.
 However,
 some new techniques have recently been developed
 for determining the local Dark Matter density
 with a higher precision
 \cite{Catena09, Weber09,
       Salucci10, Pato10, deBoer10}.
}.

 In the work on the development of
 these model--independent data analysis procedures
 for extracting WIMP properties
 from direct detection experiments,
 it was assumed that
 the analyzed data sets are background--free,
 i.e., all events are WIMP signals.
 Active background discrimination techniques
 should make this condition possible.
 For example,
 the ratio of the ionization to recoil energy,
 the so--called ``ionization yield'',
 used in the CDMS-II experiment
 provides an event--by--event rejection
 of electron recoil events
 to be better than $10^{-4}$ misidentification
 \cite{Ahmed09b}.
 By combining the ``phonon pulse timing parameter'',
 the rejection ability of
 the misidentified electron recoils
 (most of them are ``surface events''
  with sufficiently reduced ionization energies)
 can be improved to be $< 10^{-6}$ \cite{Ahmed09b}.
 Moreover,
 as demonstrated by the CRESST collaboration \cite{CRESST-bg},
 by means of inserting a scintillating foil,
 which causes some additional scintillation light
 for events induced by $\alpha$-decay of $\rmXA{Po}{210}$
 and thus shifts the pulse shapes of these events
 faster than pulses induced by WIMP interactions in the crystal,
 the pulse shape discrimination (PSD) technique
 can then easily distinguish WIMP--induced nuclear recoils
 from those induced by backgrounds%
\footnote{
 For more details
 about background discrimination techniques and status
 in currently running and projected direct detection experiments
 see e.g.,
 Refs.~\cite{Aprile09a, EDELWEISS-bg, Lang09b}.
}.

 However,
 as the most important issue in all underground experiments,
 the signal identification ability and
 possible residue background events
 which pass all discrimination criteria and
 then mix with other real WIMP--induced events in analyzed data sets
 should also be considered.
 Therefore,
 in this article,
 as a more realistic study,
 we follow our works
 on the effects of residue background events
 in direct Dark Matter detection experiments
 \cite{DMDDbg-mchi, DMDDbg-f1v}
 and want to study
 how well we could estimate
 the SI WIMP--nucleon coupling model--independently
 by using ``impure'' data sets
 and how ``dirty'' these data sets could be
 to be still useful.

 The remainder of this article is organized as follows.
 In Sec.~2
 I review the model--independent method
 for estimating the SI WIMP coupling on nucleons
 by using experimental data sets directly.
 In Sec.~3
 the effects of residue background events
 in the analyzed data sets
 on the measured energy spectrum
 as well as on the reconstructed WIMP mass
 will briefly be discussed.
 In Sec.~4
 I show numerical results of
 the reconstructed SI WIMP--nucleon coupling
 by using mixed data sets
 with different fractions of residue background events
 based on Monte Carlo simulations.
 I conclude in Sec.~5.
 Some technical details will be given in an appendix.
\section{Method for estimating the SI WIMP--nucleon coupling}
 The basic expression for the differential event rate
 for elastic WIMP--nucleus scattering is given by \cite{SUSYDM96}:
\beq
   \dRdQ
 = \calA \FQ \int_{\vmin}^{\vmax} \bfrac{f_1(v)}{v} dv
\~.
\label{eqn:dRdQ}
\eeq
 Here $R$ is the direct detection event rate,
 i.e., the number of events
 per unit time and unit mass of detector material,
 $Q$ is the energy deposited in the detector,
 $F(Q)$ is the elastic nuclear form factor,
 $f_1(v)$ is the one--dimensional velocity distribution function
 of the WIMPs impinging on the detector,
 $v$ is the absolute value of the WIMP velocity
 in the laboratory frame.
 The constant coefficient $\calA$ is defined as
\beq
        \calA
 \equiv \frac{\rho_0 \sigma_0}{2 \mchi \mrN^2}
\~,
\label{eqn:calA}
\eeq
 where $\rho_0$ is the WIMP density near the Earth
 and $\sigma_0$ is the total cross section
 ignoring the form factor suppression.
 The reduced mass $\mrN$ is defined by
\beq
        \mrN
 \equiv \frac{\mchi \mN}{\mchi + \mN}
\~,
\label{eqn:mrN}
\eeq
 where $\mchi$ is the WIMP mass and
 $\mN$ that of the target nucleus.
 Finally,
 $\vmin$ is the minimal incoming velocity of incident WIMPs
 that can deposit the energy $Q$ in the detector:
\beq
   \vmin
 = \alpha \sqrt{Q}
\~,
\label{eqn:vmin}
\eeq
 with the transformation constant
\beq
        \alpha
 \equiv \sfrac{\mN}{2 \mrN^2}
\~,
\label{eqn:alpha}
\eeq
 and $\vmax$ is the maximal WIMP velocity
 in the Earth's reference frame,
 which is related to
 the escape velocity from our Galaxy
 at the position of the Solar system,
 $\vesc~\gsim~600$ km/s.

 For spin--independent scalar WIMP interactions,
 the total cross section in Eq.~(\ref{eqn:calA})
 can be expressed as \cite{SUSYDM96,Bertone05}
\beq
   \sigmaSI
 = \afrac{4}{\pi} \mrN^2 \bBig{Z f_{\rm p} + (A - Z) f_{\rm n}}^2
\~.
\label{eqn:sigma0_scalar}
\eeq
 Here $\mrN$ is the reduced mass defined in Eq.~(\ref{eqn:mrN}),
 $Z$ is the atomic number of the target nucleus,
 i.e., the number of protons,
 $A$ is the atomic mass number,
 $A-Z$ is then the number of neutrons,
 $f_{\rm (p, n)}$ are the effective
 scalar couplings of WIMPs on protons p and on neutrons n,
 respectively.
 Here we have to sum over the couplings
 on each nucleon before squaring
 because the wavelength associated with the momentum transfer
 is comparable to or larger than the size of the nucleus,
 the so--called ``coherence effect''.
 In addition,
 for the lightest supersymmetric neutralino,
 and for all WIMPs which interact primarily through Higgs exchange,
 the scalar couplings are approximately the same
 on protons and on neutrons:
\( 
        f_{\rm n}
 \simeq f_{\rm p}
\).
 Thus the ``pointlike'' cross section $\sigmaSI$
 in Eq.~(\ref{eqn:sigma0_scalar}) can be written as
\beq
        \sigmaSI
 \simeq \afrac{4}{\pi} \mrN^2 A^2 |f_{\rm p}|^2
 =      A^2 \afrac{\mrN}{\mrp}^2 \sigmapSI
\~,
\label{eqn:sigma0SI}
\eeq
 where $\mrp$ is the reduced mass
 of the WIMP mass $\mchi$ and the proton mass $m_{\rm p}$,
 and
\beq
   \sigmapSI
 = \afrac{4}{\pi} \mrp^2 |f_{\rm p}|^2
\label{eqn:sigmapSI}
\eeq
 is the SI WIMP--nucleon cross section.
 Here the tiny mass difference between a proton and a neutron
 has been neglected.

 It was found that,
 by using a time--averaged recoil spectrum,
 and assuming that no directional information exists,
 the normalized one--dimensional
 velocity distribution function of halo WIMPs, $f_1(v)$,
 can be solved from Eq.~(\ref{eqn:dRdQ}) analytically \cite{DMDDf1v}
 and,
 consequently,
 its generalized moments can be estimated by
 \cite{DMDDf1v, DMDDmchi}%
\footnote{
 Here we have implicitly assumed that
 $\Qmax$ is so large that
 a term $2 \Qmax^{(n+1)/2} r(\Qmax) / F^2(\Qmax)$
 is negligible.
}
\beqn
    \expv{v^n}(v(\Qmin), v(\Qmax))
 \= \int_{v(\Qmin)}^{v(\Qmax)} v^n f_1(v) \~ dv
    \non\\
 \= \alpha^n
    \bfrac{2 \Qmin^{(n+1)/2} r(\Qmin) / \FQmin + (n+1) I_n(\Qmin, \Qmax)}
          {2 \Qmin^{   1 /2} r(\Qmin) / \FQmin +       I_0(\Qmin, \Qmax)}
\~.
\label{eqn:moments}
\eeqn
 Here $v(Q) = \alpha \sqrt{Q}$,
 $Q_{\rm (min, max)}$ are
 the experimental minimal and maximal
 cut--off energies of the data set,
 respectively,
\beq
        r(\Qmin)
 \equiv \adRdQ_{{\rm expt},\~Q = \Qmin}
\label{eqn:rmin}
\eeq
 is an estimated value of
 the {\em measured} recoil spectrum $(dR / dQ)_{\rm expt}$
 ({\em before} normalized by an experimental exposure $\cal E$)
 at $Q = \Qmin$,
 and $I_n(\Qmin, \Qmax)$ can be estimated through the sum:
\beq
   I_n(\Qmin, \Qmax)
 = \sum_{a = 1}^{N_{\rm tot}} \frac{Q_a^{(n-1)/2}}{F^2(Q_a)}
\~,
\label{eqn:In_sum}
\eeq
 where the sum runs over all events in the data set
 that satisfy $Q_a \in [\Qmin, \Qmax]$
 and $N_{\rm tot}$ is the number of such events.
 Note that,
 firstly,
 by using the second line of Eq.~(\ref{eqn:moments})
 $\expv{v^n}(v(\Qmin), v(\Qmax))$ can be determined
 independently of the local WIMP density $\rho_0$,
 of the velocity distribution function of incident WIMPs, $f_1(v)$,
 as well as of the WIMP--nucleus cross section $\sigma_0$.
 Secondly,
% as shown later,
 $r(\Qmin)$ and $I_n(\Qmin, \Qmax)$
 are two key quantities for our analysis,
 which can be estimated
 either from a functional form of the recoil spectrum
 or from experimental data (i.e., the measured recoil energies) directly%
\footnote{
 All formulae needed for estimating
 $r(\Qmin)$, $I_n(\Qmin, \Qmax)$, and their statistical errors
 are given in the appendix.
% can be found in Refs.~\cite{DMDDmchi, DMDDmchi-NJP}.
}.

 By substituting the first expression in Eq.~(\ref{eqn:sigma0SI})
 into Eq.~(\ref{eqn:dRdQ}),
 and using the fact that
 the integral over the one--dimensional WIMP velocity distribution
 on the right--hand side of Eq.~(\ref{eqn:dRdQ})
 is the minus--first moment of this distribution,
 which can be estimated by Eq.~(\ref{eqn:moments}) with $n = -1$,
 we have
\beqn
    \adRdQ_{{\rm expt}, \~ Q = \Qmin}
 \= \calE \calA \FQmin \int_{v(\Qmin)}^{v(\Qmax)} \bfrac{f_1(v)}{v} dv
    \non\\
 \= \calE \afrac{2 \rho_0 A^2 |f_{\rm p}|^2}{\pi \mchi} \FQmin \cdot
    \frac{1}{\alpha}
    \bfrac{2 r(\Qmin) / \FQmin}{2 \Qmin^{1/2} r(\Qmin) / \FQmin + I_0}
\~.
\label{eqn:dRdQ_expt_Qmin}
\eeqn
 Using the definition (\ref{eqn:alpha}) of $\alpha$,
 the {\em squared} SI WIMP coupling on protons (nucleons)
 can then be expressed as \cite{DMDDfp2-IDM2008, DMDDfp2}
\beq
   |f_{\rm p}|^2
 = \frac{1}{\rho_0}
   \bbrac{\frac{\pi}{4 \sqrt{2}} \afrac{1}{\calE A^2 \sqrt{\mN}}}
   \bbrac{\frac{2 \Qmin^{1/2} r(\Qmin)}{\FQmin} + I_0}
   \abrac{\mchi + \mN}
\~.
\label{eqn:fp2}
\eeq
 Note that
 the experimental exposure $\calE$
 appearing in the denominator
 relates the {\em actual} counting rate $(dR / dQ)_{\rm expt}$
 to the normalized rate in Eq.~(\ref{eqn:dRdQ}).

 As mentioned in the introduction,
 the Dark Matter density at the position of the Solar system,
 $\rho_0$,
 appearing in the denominator of the expression (\ref{eqn:fp2})
 for estimating $|f_{\rm p}|^2$
 has conventionally been estimated
 by means of the measurement of
 the rotation curve of the Milky Way.
 The currently most commonly used value for $\rho_0$ is
 \cite{SUSYDM96, Bertone05}
\beq
         \rho_0
 \approx 0.3~{\rm GeV/cm^3}
\~,
\label{eqn:rho0}
\eeq
 with an uncertainty of a factor of $\sim$ 2.
 However,
 some new techniques have been developed
 for determining $\rho_0$ with a higher precision
 \cite{Catena09, Weber09,
       Salucci10, Pato10, deBoer10}.
 These estimates give rather {\em larger} values for $\rho_0$;
 e.g., Catena and Ullio gave \cite{Catena09}
\beq
   \rho_0
 = 0.39 \pm 0.03~{\rm GeV/cm^3}
\~,
\eeq
 and Salucci {\it et al.} even gave \cite{Salucci10}
\beq
   \rho_0
 = 0.43 \pm 0.11 \pm 0.10~{\rm GeV/cm^3}
\~.
\eeq
 Moreover,
 instead of a spherical symmetric density profile
 assumed in Refs.~\cite{Catena09, Salucci10},
 in Refs.~\cite{Weber09, Pato10, deBoer10}
 the authors considered an axisymmetric density profile
 for a flattened Galactic Dark Matter halo \cite{Sackett94}
 caused by the disk structure of the luminous baryonic component.
 It was found that
 the local density of such a non--spherical DM halo
 could be enhanced by $\sim$ 20\% or larger
 \cite{Weber09, Pato10}
 and Pato {\it et al.} gave therefore \cite{Pato10}
\beq
   \rho_0
 = 0.466 \pm 0.033 ({\rm stat}) \pm 0.077 ({\rm syst})~{\rm GeV/cm^3}
\~.
\eeq
 Nevertheless,
 since the squared SI WIMP--nucleon coupling $|f_{\rm p}|^2$
 is inversely proportional to the local WIMP density,
 by using Eq.~(\ref{eqn:fp2})
 one can at least give an upper bound on $|f_{\rm p}|^2$.
 Moreover,
 as shown in Refs.~\cite{DMDDfp2-IDM2008, DMDDfp2},
 in spite of the very few ($\cal O$(50)) events
 from one experiment,
 for a WIMP mass of 100 GeV,
 the SI WIMP--nucleon coupling $|f_{\rm p}|$ can be estimated
 with a statistical uncertainty of only $\sim$ 15\%;
 it leads to an uncertainty on
 the SI WIMP--nucleon cross section of $\sim$ 30\%,
 which is (much) smaller than the uncertainty
 on the estimate of the local Dark Matter density.
\section{Effects of residue background events}
 In this section
 I first show some numerical results of
 the energy spectrum of WIMP recoil signals
 mixed with a few background events.
 Then
 I review the effects of residue background events
 in the analyzed data sets
 on the reconstruction of the WIMP mass $\mchi$.

 For generating WIMP--induced signals,
 we use the shifted Maxwellian velocity distribution
 \cite{Lewin96, SUSYDM96, DMDDf1v}:
\beq
   f_{1, \sh}(v)
 = \frac{1}{\sqrt{\pi}} \afrac{v}{\ve v_0}
   \bbigg{ e^{-(v - \ve)^2 / v_0^2} - e^{-(v + \ve)^2 / v_0^2} }
\~,
\label{eqn:f1v_sh}
\eeq
%
% Here
 with $v_0 \simeq 220~{\rm km/s}$
 and $\ve = 1.05 \~ v_0$,
 which are the Sun's orbital velocity
 and the Earth's velocity in the Galactic frame%
\footnote{
 The time dependence of the Earth's velocity
 will be ignored in our simulations.
},
 respectively;
 the maximal cut--off
 of the velocity distribution function
 has been set as $\vmax = 700$ km/s.
 The commonly used elastic nuclear form factor
 for the SI cross section
 \cite{Engel91, SUSYDM96, Bertone05}:
\beq
   F_{\rm SI}^2(Q)
 = \bfrac{3 j_1(q R_1)}{q R_1}^2 e^{-(q s)^2}
%\~.
\label{eqn:FQ_WS}
\eeq
 will also be used%
\footnote{
 Other commonly used analytic forms
 for the one--dimensional WIMP velocity distribution
 as well as for the elastic nuclear form factor
 for the SI WIMP--nucleus cross section
 can be found in Refs.~\cite{DMDDf1v, DMDDmchi-NJP}.
}.
 Meanwhile,
 in order to check
 the need of a prior knowledge about
 an (exact) form of the residue background spectrum,
 two forms for the background spectrum
 have been considered.
 The simplest choice is a constant spectrum:
\beq
   \adRdQ_{\rm bg, const}
 = 1
\~.
\label{eqn:dRdQ_bg_const}
\eeq
 More realistically,
 we use the target--dependent exponential form
 introduced in Ref.~\cite{DMDDbg-mchi}
 for the residue background spectrum:
\beq
   \adRdQ_{\rm bg, ex}
 = \exp\abrac{-\frac{Q /{\rm keV}}{A^{0.6}}}
\~.
\label{eqn:dRdQ_bg_ex}
\eeq
 Here $Q$ is the recoil energy,
 $A$ is the atomic mass number of the target nucleus.
 The power index of $A$, 0.6, is an empirical constant,
 which has been chosen so that
 the exponential background spectrum is
 somehow {\em similar to},
 but still {\em different from}
 the expected recoil spectrum of the target nucleus;
 otherwise,
 there is in practice no difference between
 the WIMP scattering and background spectra.
 Note that,
 among different possible choices,
 we use in our simulations the atomic mass number $A$
 as the simplest, unique characteristic parameter
 in the general analytic form (\ref{eqn:dRdQ_bg_ex})
 for defining the residue background spectrum
 for {\em different} target nuclei.
 However,
 it does {\em not} mean that
 the (superposition of the real) background spectra
 would depend simply/primarily on $A$ or
 on the mass of the target nucleus, $\mN$.
 In other words,
 it is practically equivalent to
 use expression (\ref{eqn:dRdQ_bg_ex})
 or $(dR / dQ)_{\rm bg, ex} = e^{-Q / 13.5~{\rm keV}}$ directly
 for a $\rmXA{Ge}{76}$ target.

 Note also that,
 firstly,
 as argued in Ref.~\cite{DMDDbg-mchi},
 two forms of background spectrum given above
 are rather naive;
 however,
 since we consider here
 only {\em a few residue} background events
 induced by perhaps {\em two or more} different sources,
 which pass all discrimination criteria,
 and then mix with other WIMP--induced events
 in our data sets of ${\cal O}(50)$ {\em total} events,
 exact forms of different background spectra
 are actually not very important and
 these two spectra,
 in particular,
 the exponential one,
 should practically not be unrealistic%
\footnote{
 Other (more realistic) forms for background spectrum
 (perhaps also for some specified targets/experiments)
 can be tested on the \amidas\ website
 \cite{AMIDAS-web, AMIDAS-eprints}.
}.
 Secondly,
 as demonstrated in Refs.~\cite{DMDDfp2-IDM2008, DMDDfp2}
 and reviewed in the previous section,
 the model--independent data analysis procedure
 for estimating the SI WIMP--nucleon coupling
 requires only measured recoil energies
 (induced mostly by WIMPs and
  occasionally by background sources)
 from direct detection experiments.
 Therefore,
 for applying this method to future real data,
 a prior knowledge about (different) background source(s)
 is {\em not required at all}.

 Moreover,
 for our numerical simulations
 presented here as well as in the next section,
 the actual numbers of signal and background events
 in each simulated experiment
 are Poisson--distributed around their expectation values
 {\em independently};
 and the total event number recorded in one experiment
 is then the sum of these two numbers.
 Additionally,
 we assumed that
 all experimental systematic uncertainties
 as well as the uncertainty on
 the measurement of the recoil energy
 could be ignored.
 The energy resolution of most existing detectors
 is so good that its error can be neglected
 compared to the statistical uncertainty
 for the foreseeable future
 with pretty few events.

\begin{figure}[p!]
\begin{center}
\vspace{-0.75cm}
\hspace*{-1.6cm}
\includegraphics[width=9.8cm]{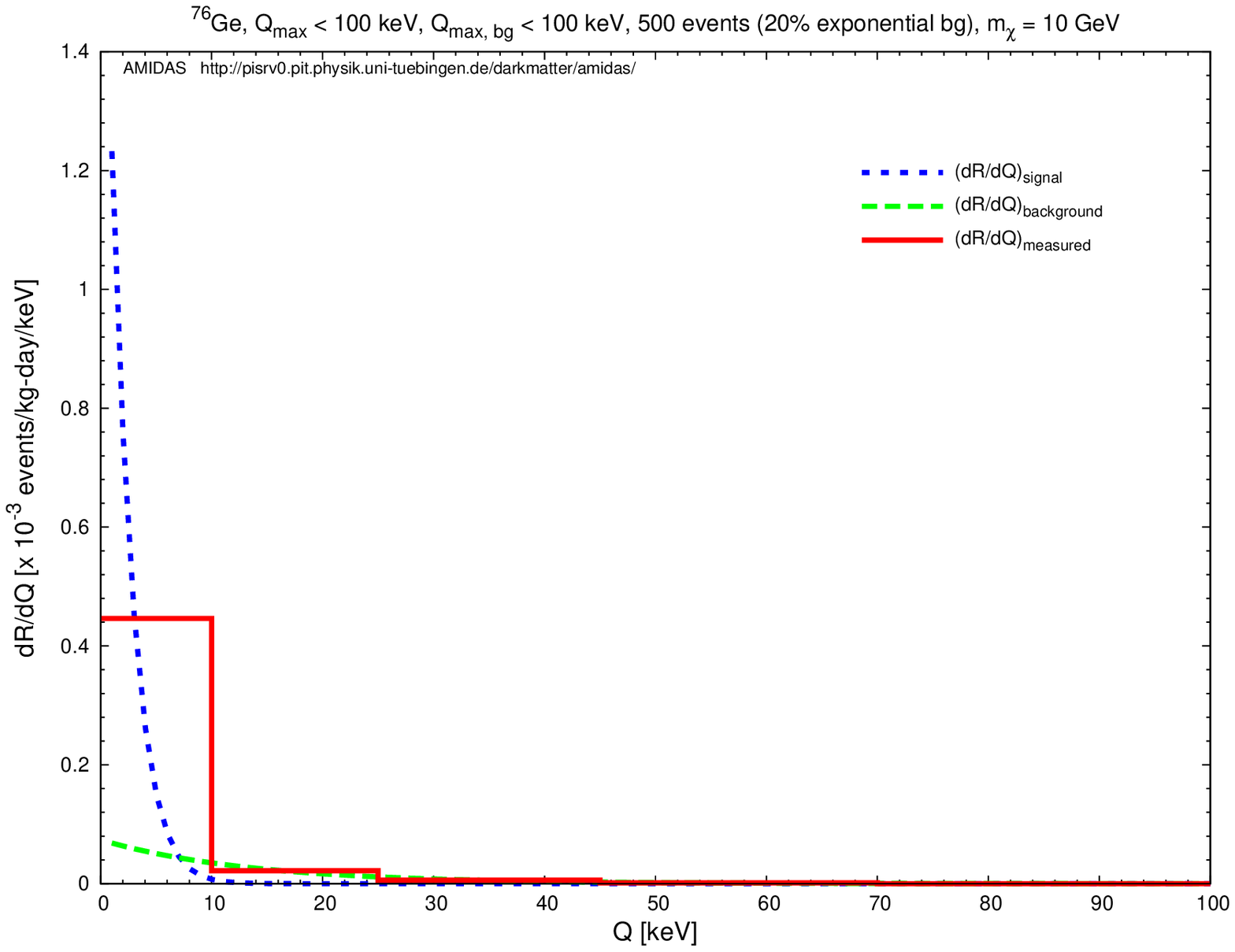} \hspace{-1.1cm}
\includegraphics[width=9.8cm]{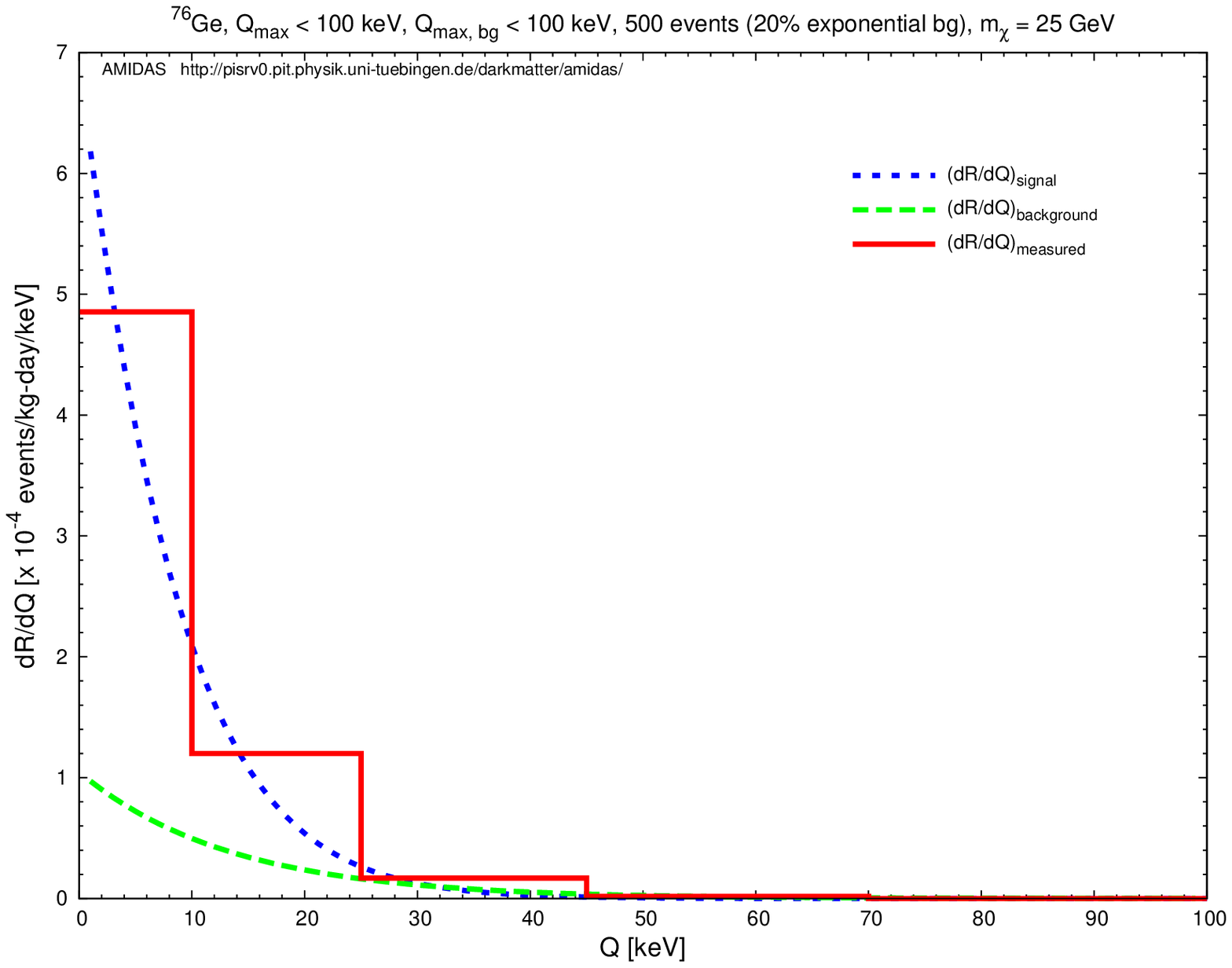} \hspace*{-1.6cm} \\
\vspace{0.5cm}
\hspace*{-1.6cm}
\includegraphics[width=9.8cm]{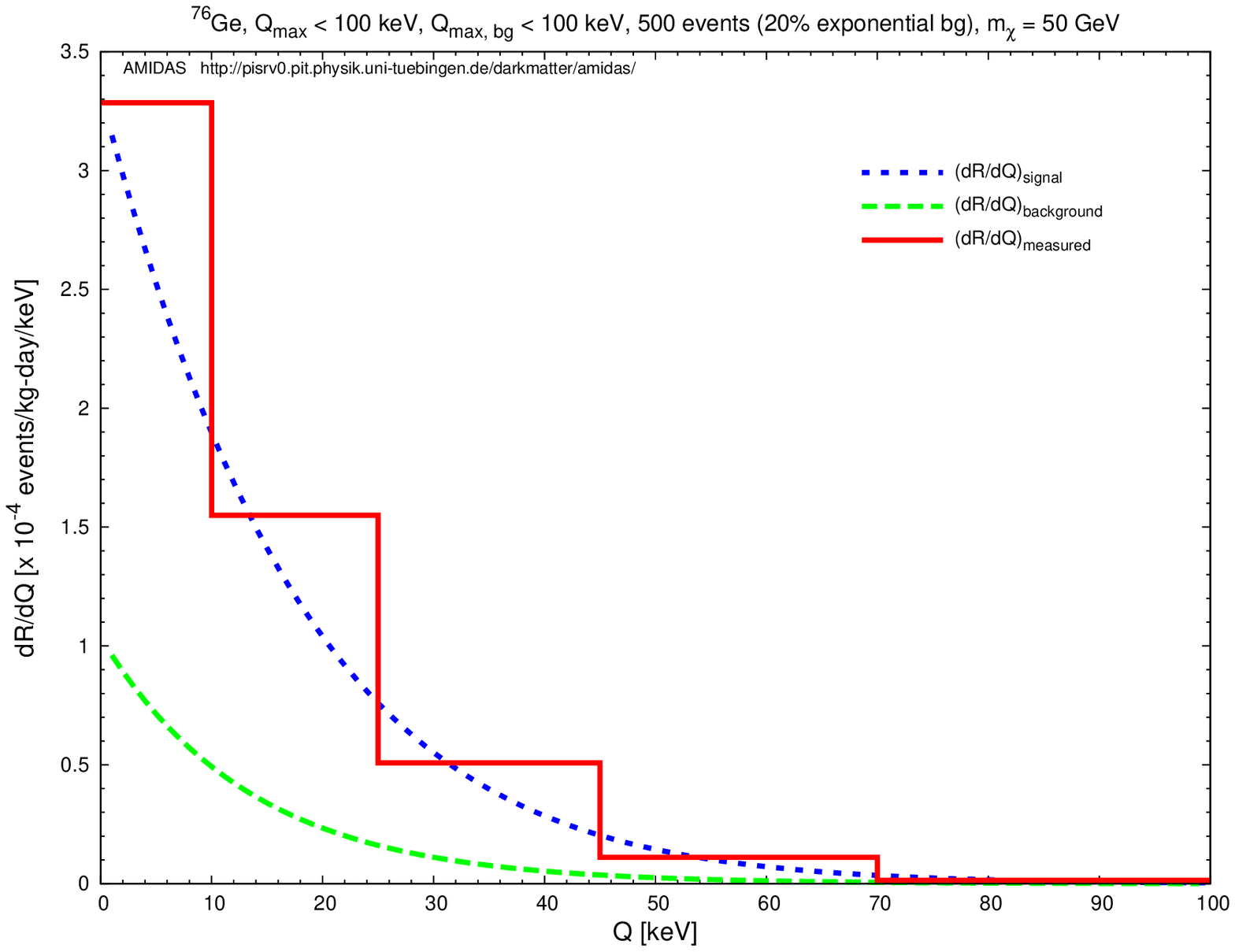} \hspace{-1.1cm}
\includegraphics[width=9.8cm]{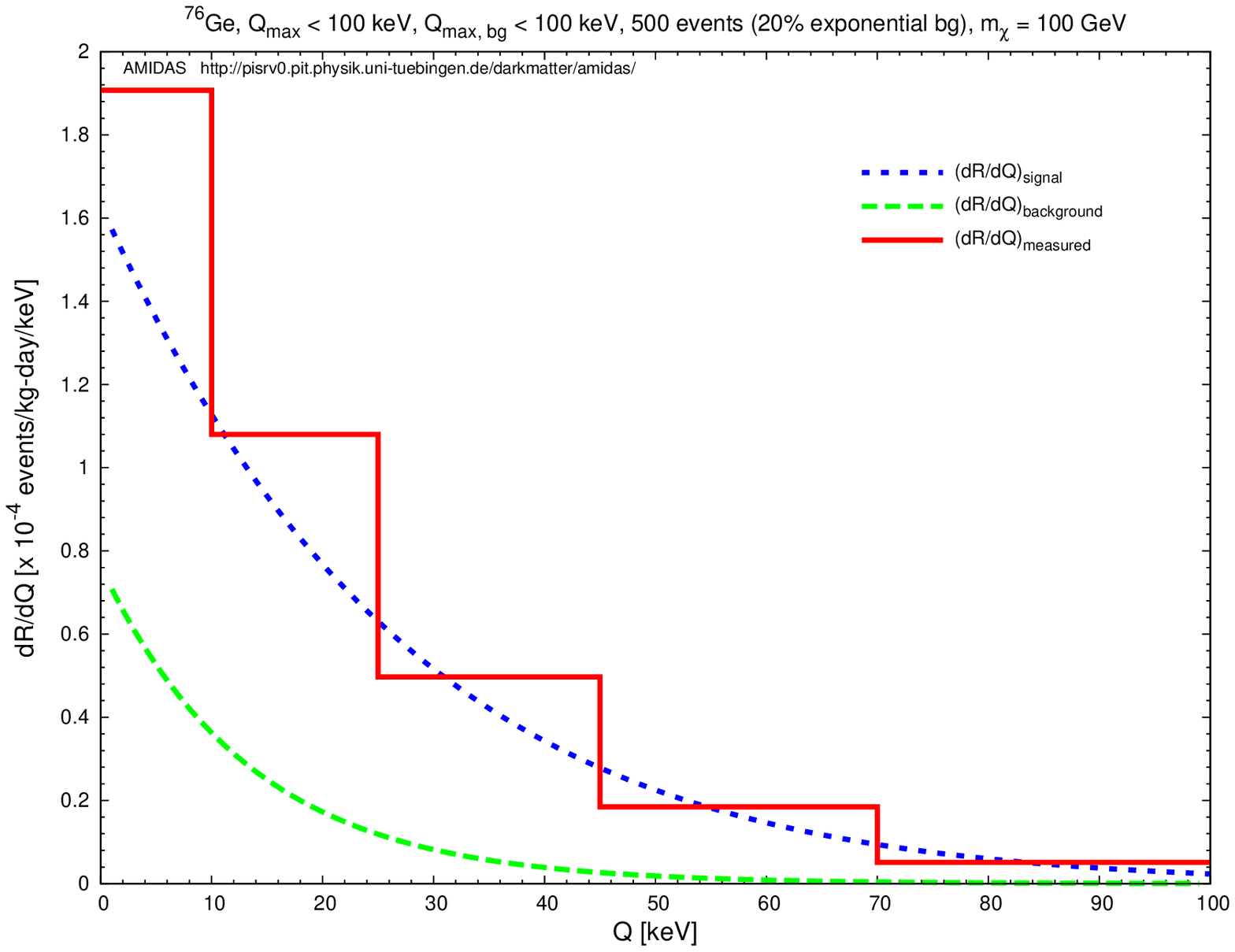} \hspace*{-1.6cm} \\
\vspace{0.5cm}
\hspace*{-1.6cm}
\includegraphics[width=9.8cm]{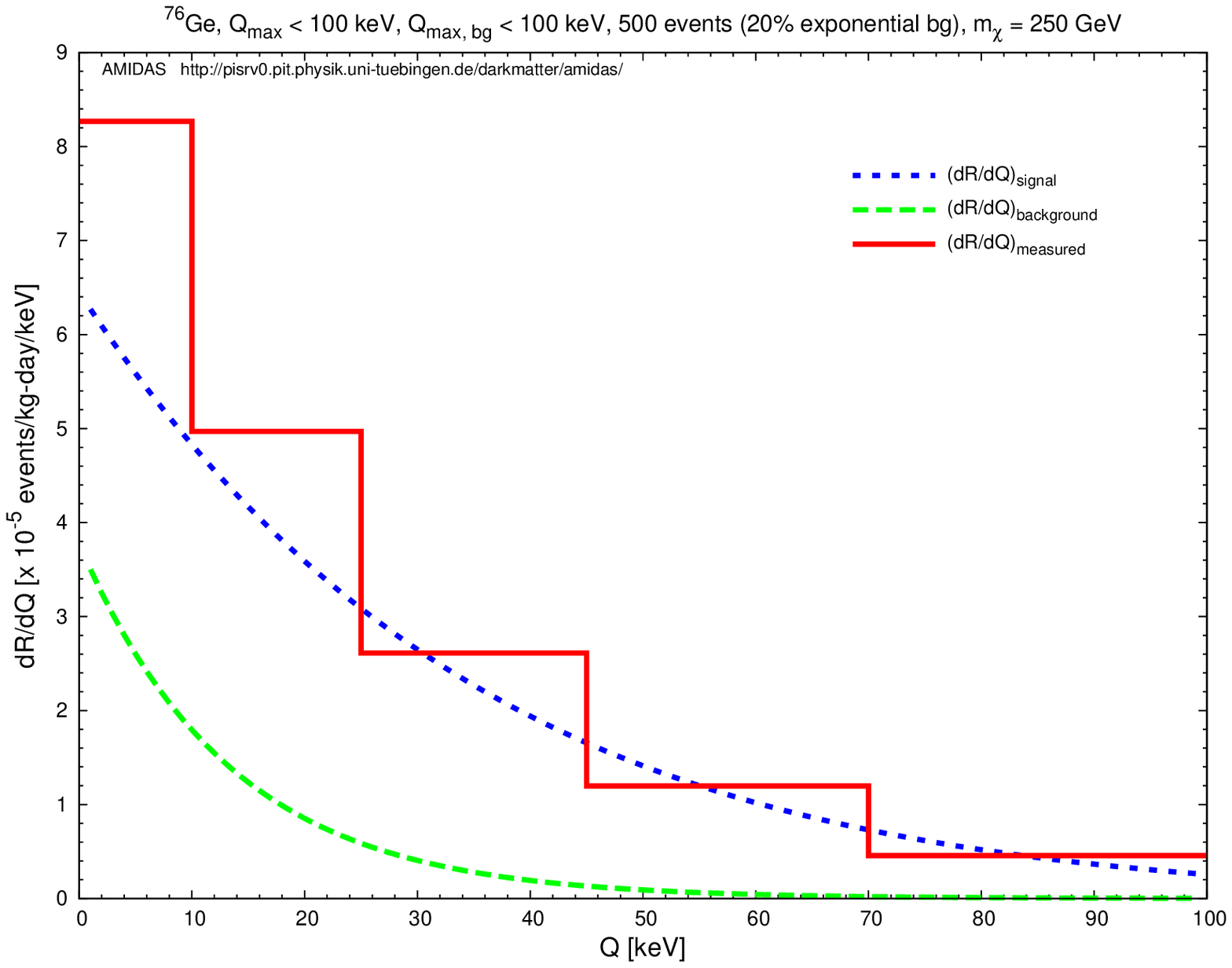} \hspace{-1.1cm}
\includegraphics[width=9.8cm]{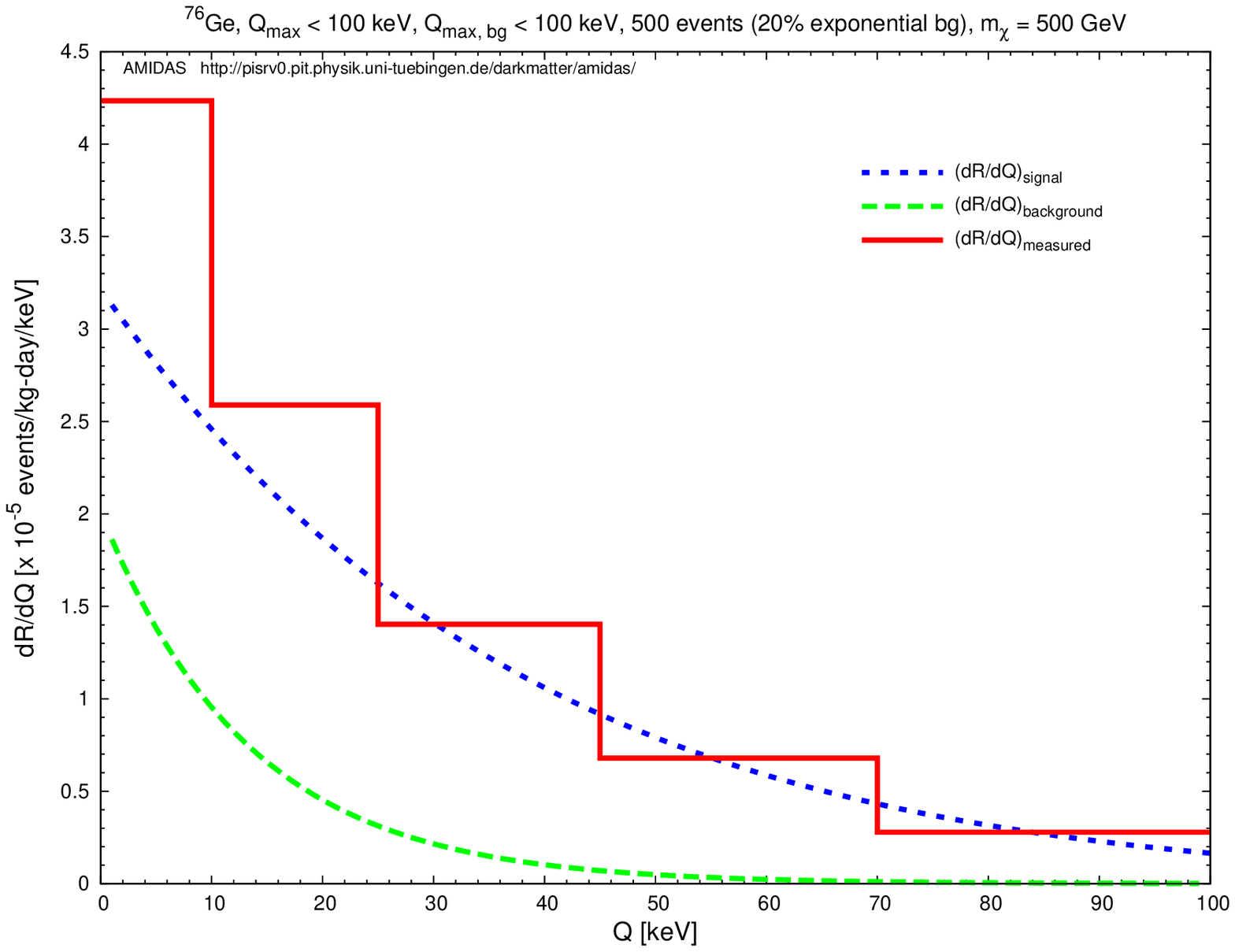} \hspace*{-1.6cm} \\
\vspace{-0.75cm}
\end{center}
\caption{
 Measured energy spectra (solid red histograms)
 for a $\rmXA{Ge}{76}$ target
 with six different WIMP masses:
 10, 25, 50, 100, 250, and 500 GeV.
 The dotted blue curves are
 the elastic WIMP--nucleus scattering spectra,
 whereas
 the dashed green curves are
 the exponential background spectra
 normalized to fit to the chosen background ratio,
 which has been set as 20\% here.
 The experimental threshold energies
 have been assumed to be negligible
 and the maximal cut--off energies
 are set as 100 keV.
 The background windows
 have been assumed to be the same as
 the experimental possible energy ranges.
 5,000 experiments with 500 total events on average
 in each experiment have been simulated.
 See the text for further details
 (plots from Ref.~\cite{DMDDbg-mchi}).
}
\label{fig:dRdQ-bg-ex-Ge-000-100-20}
\end{figure}
\begin{figure}[p!]
\begin{center}
\vspace{-1cm}
\includegraphics[width=15cm]{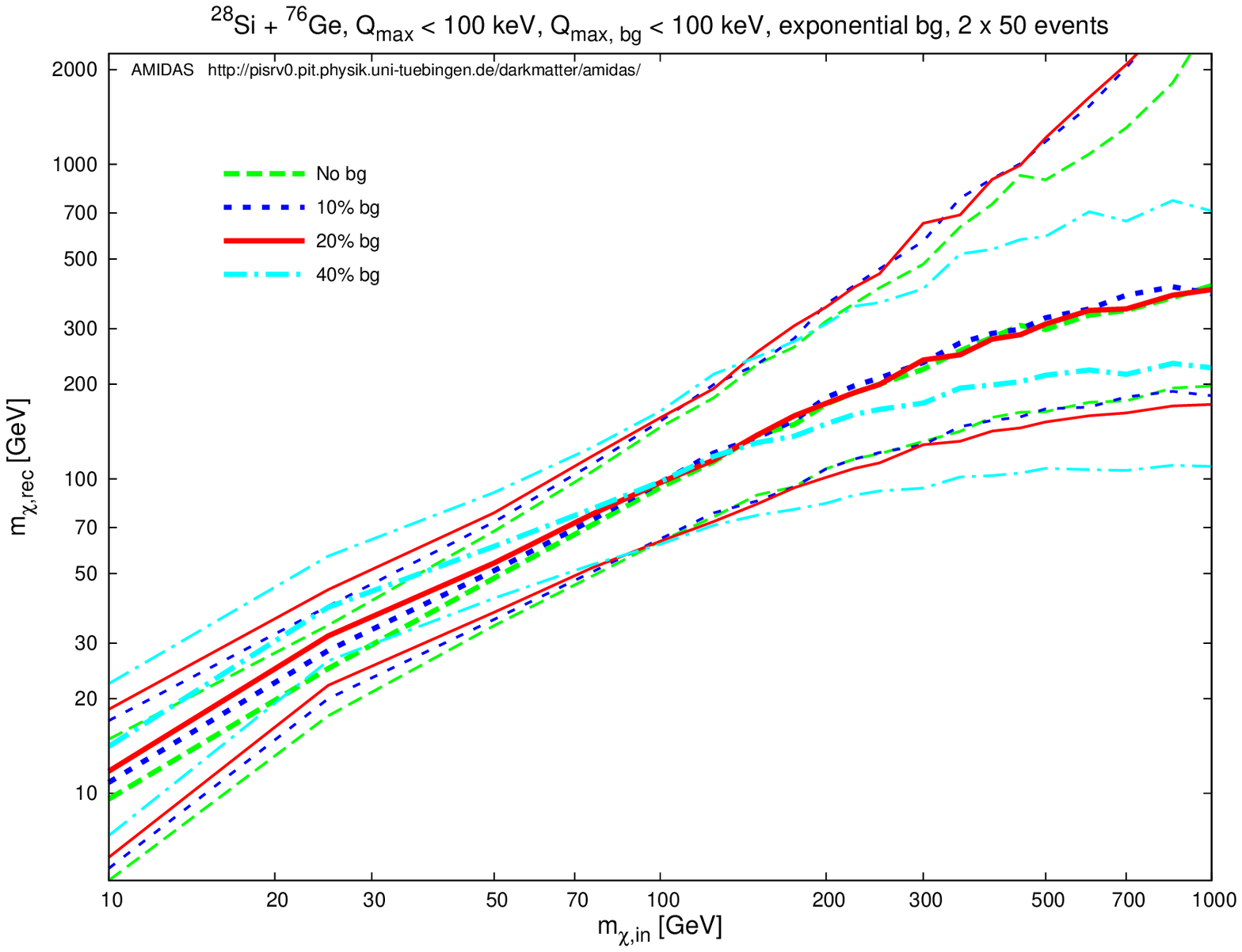}    \\ \vspace{0.2cm}
\includegraphics[width=15cm]{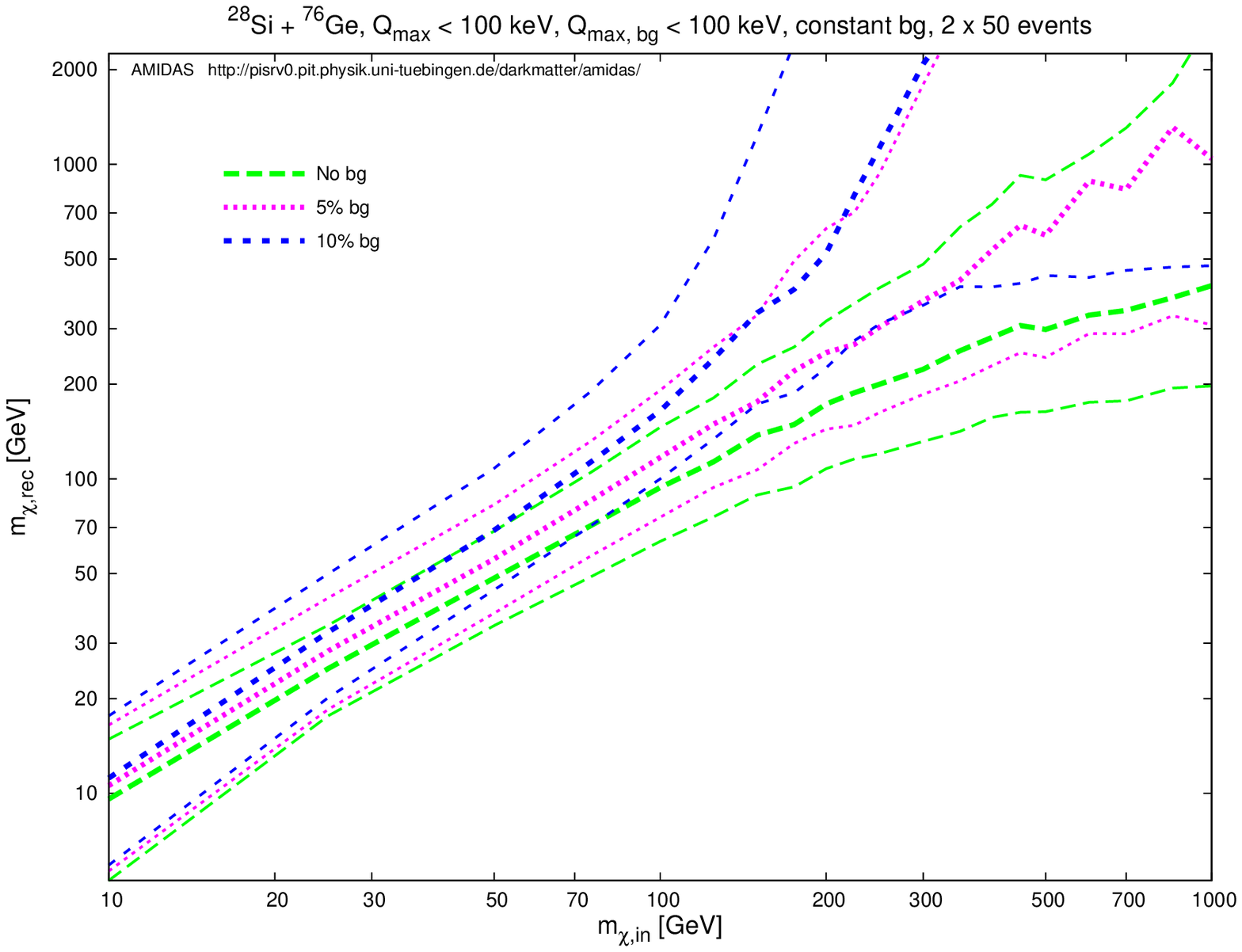} \\
\vspace{-0.55cm}
\end{center}
\caption{
 The reconstructed WIMP masses
 and the lower and upper bounds of
 the 1$\sigma$ statistical uncertainties
 by using mixed data sets
 from WIMP--induced and background events
 as functions of the input WIMP mass.
 $\rmXA{Si}{28}$ and $\rmXA{Ge}{76}$
 have been chosen as two target nuclei.
 The exponential (upper) and
 constant (lower) forms given
 in Eqs.~(\ref{eqn:dRdQ_bg_ex}) and (\ref{eqn:dRdQ_bg_const})
 have been used for the background spectrum.
 The background ratios shown here
 are no background (dashed green),
  5\% (dotted magenta),
 10\% (long--dotted blue),
 20\% (solid red),
 and 40\% (dash--dotted cyan)
 background events in the analyzed data sets
 in the experimental energy ranges
 between 0 and 100 keV.
 Each experiment contains 50 total events
 on average.
 Other parameters are as
 in Figs.~\ref{fig:dRdQ-bg-ex-Ge-000-100-20}.
 See the text for further details
 (plots from Ref.~\cite{DMDDbg-mchi}).
}
\label{fig:mchi-SiGe-000-100-050}
\end{figure}
\subsection{On the measured energy spectrum}

 In Figs.~\ref{fig:dRdQ-bg-ex-Ge-000-100-20}
 I show measured energy spectra (solid red histograms)
 for a $\rmXA{Ge}{76}$ target
 with six different WIMP masses:
 10, 25, 50, 100, 250, and 500 GeV
 based on Monte Carlo simulations.
 The dotted blue curves are
 the elastic WIMP--nucleus scattering spectra,
 whereas
 the dashed green curves are
 the exponential background spectra
 given in Eq.~(\ref{eqn:dRdQ_bg_ex}),
 which have been normalized so that
 the ratios of the areas under these background spectra
 to those under the (dotted blue) WIMP scattering spectra
 are equal to the background--signal ratio
 in the whole data sets
 (e.g.,~20\% backgrounds to 80\% signals
  shown in Figs.~\ref{fig:dRdQ-bg-ex-Ge-000-100-20}).
 The experimental threshold energies
 have been assumed to be negligible
 and the maximal cut--off energies
 are set as 100 keV.
 The background windows
 (the possible energy ranges
  in which residue background events exist)
 have been assumed to be the same as
 the experimental possible energy ranges.
 5,000 experiments with 500 total events on average
 in each experiment have been simulated.

 Remind that
 the measured energy spectra shown here
 are averaged over the simulated experiments.
 Five bins with linear increased bin widths
 have been used for binning
 generated signal and background events.
 As argued in Ref.~\cite{DMDDf1v},
 for reconstructing the one--dimensional
 WIMP velocity distribution function,
 this unusual, particular binning has been chosen
 in order to accumulate more events
 in high energy ranges
 and thus to reduce the statistical uncertainties
 in high velocity ranges.
 However,
 as shown in Sec.~2,
 for the estimation of the SI WIMP--nucleon coupling
 (as well as for the determination of the WIMP mass \cite{DMDDmchi}),
 one needs either events in the {\em first} energy bin
 or {\em all} events in the whole data set.
 Hence,
 there is in practice no difference
 between using an equal bin width for all bins
 or a (linear) increased bin widths.

 It can be found
 in Figs.~\ref{fig:dRdQ-bg-ex-Ge-000-100-20} that
 the shape of the WIMP scattering spectrum
 depends highly on the WIMP mass:
 for light WIMPs ($\mchi~\lsim~50$ GeV),
 the recoil spectra drop sharply with increasing recoil energies,
 while for heavy WIMPs ($\mchi~\gsim~100$ GeV),
 the spectra become flatter.
 In contrast,
 the exponential background spectra shown here
 depend only on the target mass
 and are rather flatter (sharper)
 for light (heavy) WIMP masses
 compared to the WIMP scattering spectra.
 This means that,
 once input WIMPs are light (heavy),
 background events would contribute relatively more to
 high (low) energy ranges,
 and, consequently,
 the measured energy spectra
 would mimic scattering spectra
 induced by heavier (lighter) WIMPs.
 Moreover,
 for heavy WIMP masses,
 since background events would contribute relatively more to
 low energy ranges,
 the estimated value of the measured recoil spectrum
 at the lowest experimental cut--off energy, $r(\Qmin)$,
 could thus be (strongly) overestimated.

 More detailed illustrations and discussions
 about the effects of residue background events
 on the measured energy spectrum
 can be found in Ref.~\cite{DMDDbg-mchi}.
\subsection{On the reconstructed WIMP mass}

 Figs.~\ref{fig:mchi-SiGe-000-100-050}
 show the reconstructed WIMP masses
 and the lower and upper bounds of
 the 1$\sigma$ statistical uncertainties
 by means of the model--independent procedure introduced
 in Refs.~\cite{DMDDmchi-SUSY07, DMDDmchi}
 with mixed data sets
 from WIMP--induced and background events
 as functions of the input WIMP mass.
 As in Ref.~\cite{DMDDmchi},
 $\rmXA{Si}{28}$ and $\rmXA{Ge}{76}$
 have been chosen as two target nuclei.
 The experimental threshold energies of two experiments
 have been assumed to be negligible
 and the maximal cut--off energies
 are set the same as 100 keV.
 The exponential (upper) and
 constant (lower) forms given
 in Eqs.~(\ref{eqn:dRdQ_bg_ex}) and (\ref{eqn:dRdQ_bg_const})
 have been used for the background spectrum.
 The background windows are set
 the same as the experimental possible energy ranges
 for both experiments.
 The background ratios shown here
 are no background (dashed green),
  5\% (dotted magenta),
 10\% (long--dotted blue),
 20\% (solid red),
 and 40\% (dash--dotted cyan)
 background events in the analyzed data sets.
 2 $\times$ 5,000 experiments have been simulated.
 Each experiment contains 50 {\em total} events
 on average.
 Note that
 {\em all} events recorded in our data sets
 are treated as WIMP signals in the analysis,
 although statistically we know that
 a fraction of these events could be backgrounds.

 From the upper frame of Figs.~\ref{fig:mchi-SiGe-000-100-050}
 it can be seen clearly that,
 for light WIMP masses ($\mchi~\lsim~100$ GeV),
 caused by the relatively flatter background spectrum
 (compared to the scattering spectrum induced by light WIMPs),
 the energy spectrum of all recorded events
 would mimic a scattering spectrum induced
 by WIMPs with a relatively heavier mass,
 and, consequently,
 the reconstructed WIMP masses
 as well as the statistical uncertainty intervals
 could be {\em overestimated}.
 In contrast,
 for heavy WIMP masses ($\mchi~\gsim~100$ GeV),
 caused by the relatively sharper background spectrum,
 relatively more background events
 contribute to low energy ranges,
 and the energy spectrum of all recorded events
 would mimic a scattering spectrum induced
 by WIMPs with a relatively lighter mass.
 Hence,
 the reconstructed WIMP masses
 as well as the statistical uncertainty intervals
 could be {\em underestimated}.

 As a comparison,
 the lower frame of Figs.~\ref{fig:mchi-SiGe-000-100-050}
 shows that,
 since the constant background spectrum
 is flatter for all WIMP masses%
\footnote{
 Illustrations and detailed discussions about
 the effects of the constant form of
 the residue background spectrum
 on the measured energy spectrum
 for different input WIMP masses
 can be found in Ref.~\cite{DMDDbg-mchi}.
},
 background events contribute always relatively more to
 high energy ranges,
 and the measured energy spectra
 would thus always mimic scattering spectra
 induced by heavier WIMPs.
 Therefore,
 the reconstructed WIMP masses
 as well as the statistical uncertainty intervals
 are {\em overestimated} for {\em all} input WIMP masses.

 Moreover,
 Figs.~\ref{fig:mchi-SiGe-000-100-050} show that
 the larger the fraction of background events
 in the analyzed data sets,
 the more strongly over-/underestimated
 the reconstructed WIMP masses
 as well as the statistical uncertainty intervals.
 Nevertheless,
 it can be found that,
 with $\sim$ 10\% -- 20\% residue background events
 in the analyzed data sets
 of $\sim$ 50 total events,
 the 1$\sigma$ statistical uncertainty band
 could in principle cover the true WIMP mass pretty well.

 More detailed illustrations and discussions
 about the effects of residue background events
 on the determination of the WIMP mass
 can be found in Ref.~\cite{DMDDbg-mchi}.
\section{Results of the reconstructed SI WIMP--nucleon coupling}
 In this section
 I present simulation results
 of the reconstructed SI WIMP coupling on nucleons%
\footnote{
 Note that,
 rather than the mean values,
 the (bounds on the) reconstructed $|f_{\rm p}|^2$
 are always the median values
 of the simulated results.
}
 by means of the model--independent method
 described in Sec.~2
 with mixed data sets
 from WIMP--induced and background events.
 The WIMP mass $\mchi$ appearing
 in the expression (\ref{eqn:fp2})
 for estimating $|f_{\rm p}|^2$
 has been assumed to be known precisely
 from other (e.g., collider) experiments
 with an overall uncertainty of 5\% of
 the input (true) WIMP mass
 or determined from {\em other}
 two direct detection experiments%
\footnote{
 As in Refs.~\cite{DMDDfp2-IDM2008, DMDDfp2},
 in order to avoid complicated calculations of the correlations
 between the uncertainty on $\mchi$ estimated by the algorithmic procedure
 and those on $r(\Qmin)$ and $I_0$,
 we assumed here that
 the two data sets using the Ge target
 are {\em independent} of each other.
}.
 The SI WIMP--nucleon cross section for our simulations
 is set as $\sigmapSI = 10^{-8}$ pb,
 the currently most commonly used value for the local WIMP density,
 $\rho_0 = 0.3~{\rm GeV/cm^3}$,
 needed in Eq.~(\ref{eqn:fp2})
 has been used for both simulations and data analyses.
 A $\rmXA{Ge}{76}$ nucleus has been chosen
 as our detector target for reconstructing $|f_{\rm p}|^2$,
 whereas a $\rmXA{Si}{28}$ target
 and a {\em second} $\rmXA{Ge}{76}$ target
 have been used for determining $\mchi$.
 The experimental threshold energies of all experiments
 have been assumed to be negligible
 and the maximal cut--off energies
 are set the same as 100 keV.
 The exponential background spectrum
 given in Eq.~(\ref{eqn:dRdQ_bg_ex})
 has been used for generating background events
 in windows of the entire experimental possible ranges.
 (3 $\times$) 5,000 experiments
 have been simulated.

\begin{figure}[t!]
\begin{center}
\includegraphics[width=15cm]{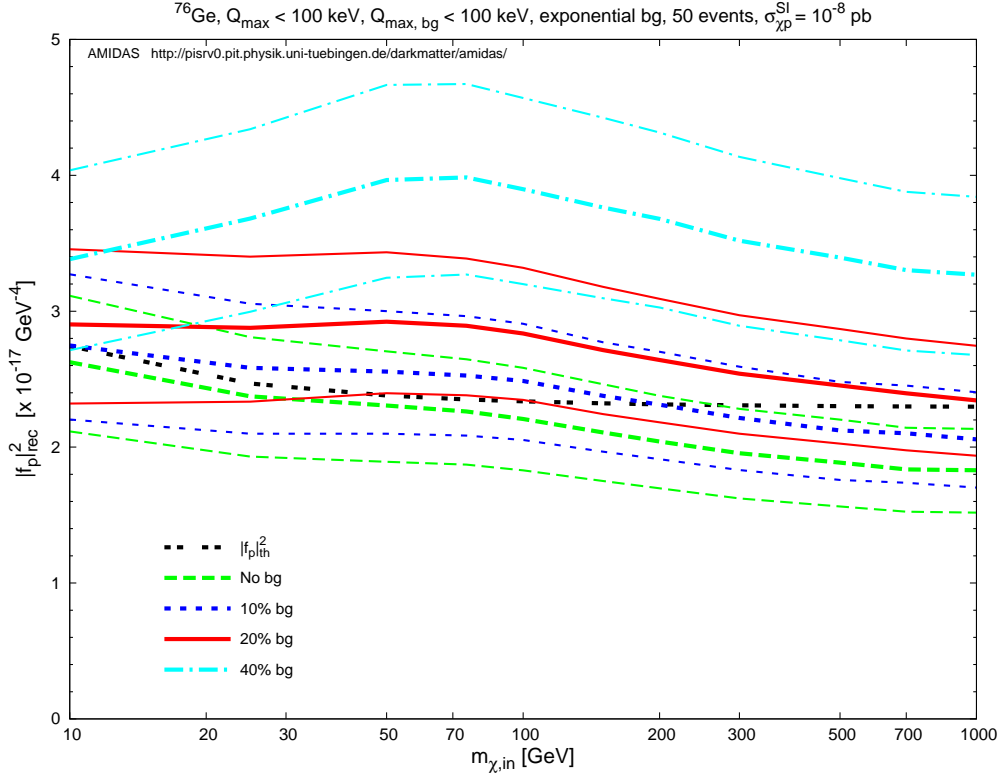} \\
\vspace{-0.5cm}
\end{center}
\caption{
 The reconstructed {\em squared}
 SI WIMP--nucleon couplings
 and the lower and upper bounds of
 the 1$\sigma$ statistical uncertainties
 by using mixed data sets
 from WIMP--induced and background events
 as functions of the input WIMP mass.
 A $\rmXA{Ge}{76}$ nucleus has been chosen
 as the target.
 The WIMP mass $\mchi$ needed in Eq.~(\ref{eqn:fp2})
 has been assumed to be known precisely.
 The background ratios shown here
 are no background (dashed green),
 10\% (long--dotted blue),
 20\% (solid red),
 and 40\% (dash--dotted cyan)
 background events in the analyzed data sets
 in the experimental energy ranges
 between 0 and 100 keV.
 The double--dotted black curve indicates
 the theoretical value of $|f_{\rm p}|^2$
 corresponding to the fixed SI WIMP--nucleon cross section
 \mbox{$\sigmapSI = 10^{-8}$ pb}.
 Each data set contains 50 total events
 on average.
 Other parameters are as
 in Figs.~\ref{fig:dRdQ-bg-ex-Ge-000-100-20}.
 See the text for further details.
}
\label{fig:fp2-Ge-ex-000-100-050}
\end{figure}

 Fig.~\ref{fig:fp2-Ge-ex-000-100-050}
 shows the reconstructed {\em squared}
 SI WIMP--nucleon couplings, $|f_{\rm p}|^2$,
 and their lower and upper bounds of
 the 1$\sigma$ statistical uncertainties
 by using mixed data sets
 as functions of the input WIMP mass.
 The WIMP mass $\mchi$ needed in Eq.~(\ref{eqn:fp2})
 has been assumed to be known precisely.
 The background ratios shown here
 are no background (dashed green),
 10\% (long--dotted blue),
 20\% (solid red),
 and 40\% (dash--dotted cyan)
 background events in the analyzed data sets
 in the experimental energy ranges
 between 0 and 100 keV.
 Each experiment contains 50 {\em total} events
 on average.
 Remind that
 {\em all} events recorded in our data sets
 are treated as WIMP signals in the analysis.

 It can be found
 in Fig.~\ref{fig:fp2-Ge-ex-000-100-050} that
 the larger the background ratio in the analyzed data set,
 the more strongly {\em overestimated}
 the reconstructed SI WIMP--nucleon coupling
 for {\em all} input WIMP masses.
 This can be understood
 from the expression (\ref{eqn:dRdQ}) for
 the different event rate $dR / dQ$.
 For a given WIMP mass and a specified target nucleus,
 the SI WIMP--nucleus cross section
 is proportional to the total event number--to--exposure ratio%
\footnote{
 Since $(dR / dQ)_{\rm expt}$ is
 the measured recoil spectrum
 before normalized by the exposure
 $\calE = \calE_{\rm sg}$,
 $R_{\rm expt}(\Qmin, \Qmax)$ here is in fact
 the total number of observed events
 $N_{\rm tot} = N_{\rm sg} + N_{\rm bg}$.
}:
\beq
         \sigmaSI
 \simeq  \afrac{4}{\pi} \mrN^2 A^2 |f_{\rm p}|^2
 \propto \frac{1}{\calE}
         \int_{\Qmin}^{\Qmax} \adRdQ_{\rm expt} dQ
 =       \frac{R_{\rm expt}(\Qmin, \Qmax)}{\calE_{\rm sg}}
\~.
%\label{eqn:}
\eeq
 Here and throughout the subscripts ``sg'' and ``bg''
 stand for WIMP signals and background events,
 respectively;
 $\calE_{\rm sg}$ is the required exposure
 to observe the expected ``WIMP signal''
 ({\em not} total) events.
 Note that,
 since background events in our data set
 are in fact {\em unexpected},
 the exposure $\calE$ in Eqs.~(\ref{eqn:dRdQ_expt_Qmin}) and (\ref{eqn:fp2})
 should thus be equal to $\calE_{\rm sg}$.
 For a fixed number of total ``observed'' events,
 the larger the background ratio,
 or, equivalently,
 the smaller the number of real WIMP--induced events,
 the smaller the required exposure $\calE = \calE_{\rm sg}$
 for accumulating the total observed events,
 and, therefore,
 the larger the estimated SI WIMP--nucleon coupling.
 In other words,
 due to {\em extra unexpected} background events
 in our data set,
 one will use a {\em larger} number of total events
 to estimate the SI WIMP coupling,
 and thus {\em overestimate} it.

 More exactly (and mathematically),
 we can separate the prefactor in the second bracket
 on the right--hand side of Eq.~(\ref{eqn:fp2})
 into two terms:
\beq
    \frac{1}{\calE}\!
    \bbrac{\frac{2 \Qmin^{1/2} r(\Qmin)}{\FQmin} + I_0\!}
 =  \frac{1}{\calE_{\rm sg}}\!
    \bbrac{\frac{2 \Qmin^{1/2} r_{\rm sg}(\Qmin)}{\FQmin} + I_{0, \rm sg}\!}
  + \frac{1}{\calE_{\rm sg}}\!
    \bbrac{\frac{2 \Qmin^{1/2} r_{\rm bg}(\Qmin)}{\FQmin} + I_{0, \rm bg}\!}
.
\label{eqn:rmin_I0}
\eeq
 It can thus be seen clearly that
 the prefactor in the expression (\ref{eqn:fp2})
 for estimating $|f_{\rm p}|^2$ would always be {\em overestimated}
 with {\em non--negligible} background events.
 Remind that
 $r(\Qmin)$ and $I_0$
 given in Eqs.~(\ref{eqn:rmin}) and (\ref{eqn:In_sum})
 are estimated from the measured recoil spectrum
 $(dR / dQ)_{\rm expt}$
 before normalized by $\calE$
 (or, equivalently, $\calE_{\rm sg}$).
 Hence,
 while the first term on the right--hand side
 of Eq.~(\ref{eqn:rmin_I0})
 remains unchanged by increasing the background ratio
 (and in turn with a decreased number of WIMP--induced events),
 the second term above
 contributed from residue background events
 causes the overestimate of
 the reconstructed SI WIMP coupling.
 Remind also that
 the experimental minimal cut--off energy, $\Qmin$,
 has been set to be negligible.
 Thus the first term involving $\Qmin^{1/2} r_{\rm bg}(\Qmin)$
 in the bracket of the second term above
 does not contribute to the reconstructed $|f_{\rm p}|^2$
 in our simulations shown here;
 otherwise,
 the reconstructed $|f_{\rm p}|^2$
 could be more strongly overestimated,
 especially for WIMP masses
 \mbox{$\mchi~\gsim~50$ GeV}
 (see Figs.~\ref{fig:dRdQ-bg-ex-Ge-000-100-20}).

 Moreover,
 for three cases with background ratios $\lsim$ 20\%
 shown in Fig.~\ref{fig:fp2-Ge-ex-000-100-050},
 the larger the {\em input WIMP mass},
 the more strongly overestimated the SI WIMP coupling.
 However,
 interestingly,
 once the background ratio rises to $\gsim$ 20\%
 (the dash--dotted cyan curves indicate
  a background ration of 40\%),
 a hump at an input WIMP mass of $\sim$ 60 GeV appears%
\footnote{
 Remind that
 the actual values of
 the ``critical'' background ratio and
 the ``critical'' WIMP mass
 (with the largest systematic deviation)
 depend in practice strongly on the WIMP scattering spectrum
 as well as on the residue background spectrum
 and therefore differ from experiment to experiment.
}.
 The reason is as follows.
 In the appendix
 I will show that
 the second term on the right--hand side
 of Eq.~(\ref{eqn:rmin_I0}) is proportional to
 the ``WIMP scattering'' spectrum
 ({\em not} the ``background'' spectrum!):
\beqn
            \frac{1}{\calE_{\rm sg}}
            \bbrac{\frac{2 \Qmin^{1/2} r_{\rm bg}(\Qmin)}{\FQmin} + I_{0, \rm bg}}
 \eqnpropto \frac{r_{\rm bg}}{1 - r_{\rm bg}}
            \int_{\Qmin}^{\Qmax} \adRdQ_{\rm sg} dQ
            \non\\
 \eqnequiv  \afrac{r_{\rm bg}}{1 - r_{\rm bg}}
            R_{\rm sg}(\Qmin, \Qmax)
\~.
%\label{eqn:}
\eeqn
 Here $(dR / dQ)_{\rm sg}$ and $R_{\rm sg}$
 are the {\em normalized} differential and total event rate
 of WIMP signals,
 respectively;
 $r_{\rm bg}$ is the ratio of residue background events
 in the whole data set.
 It can be understood from Eq.~(\ref{eqn:dRdQ}) that
 $(dR / dQ)_{\rm sg}$ and therefore $R_{\rm sg}$ are
 functions of the input (true) WIMP mass,
 through not only $\mchi$ and $\mrN$
 in the denominator of $\calA$
 defined in Eq.~(\ref{eqn:calA}),
 but also the transformation constant $\alpha$
 in Eq.~(\ref{eqn:alpha}).

\begin{figure}[t!]
\begin{center}
\includegraphics[width=12cm]{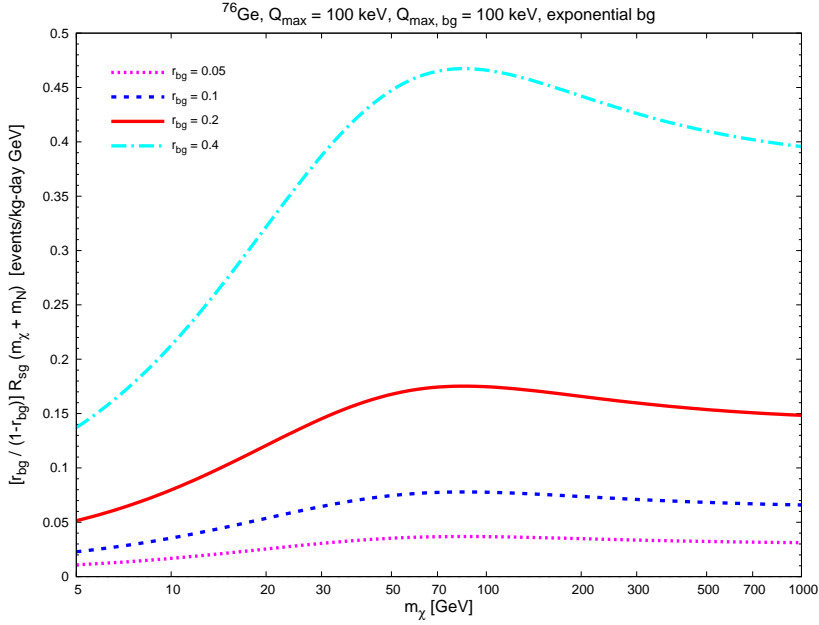} \\
\vspace{-0.5cm}
\end{center}
\caption{
 The products of
 $\bbig{r_{\rm bg} / (1 - r_{\rm bg})} \~ R_{\rm sg}$
 and $(\mchi + \mN)$
 as functions of $\mchi$
 for a $\rmXA{Ge}{76}$ target
 with different background ratios $r_{\rm bg}$.
 The background ratios shown here are
  5\% (dotted magenta),
 10\% (long--dotted blue),
 20\% (solid red),
 and 40\% (dash--dotted cyan).
 Parameters are as
 in Fig.~\ref{fig:fp2-Ge-ex-000-100-050}.
 See the text for further details.
}
\label{fig:Rsg-mchi-Ge-000-100}
\end{figure}

 Fig.~\ref{fig:Rsg-mchi-Ge-000-100} shows
 the products of
 $\bbig{r_{\rm bg} / (1 - r_{\rm bg})} \~ R_{\rm sg}$
 and $(\mchi + \mN)$
 as functions of $\mchi$
 for a $\rmXA{Ge}{76}$ target
 with different background ratios $r_{\rm bg}$.
 It can be found that
 the extra contribution from residue background events,
 which is proportional to the product of
 $\bbig{r_{\rm bg} / (1 - r_{\rm bg})} \~ R_{\rm sg}$
 and $(\mchi + \mN)$,
 has indeed a maximum at a WIMP mass of $\sim$ 75 GeV.
 Considering the slightly decreased $|f_{\rm p}|^2$ value
 by increasing the input WIMP mass
 without background events
 (the dashed green curves
  in Fig.~\ref{fig:fp2-Ge-ex-000-100-050}),
 the total recorded events
 (including WIMP--induced and background events)
 should thus result in a hump of the reconstructed $|f_{\rm p}|^2$
 at an input WIMP mass of \mbox{$\sim$ 60 GeV},
 once the background fraction in our data set
 is large enough.

\begin{figure}[t!]
\begin{center}
\includegraphics[width=15cm]{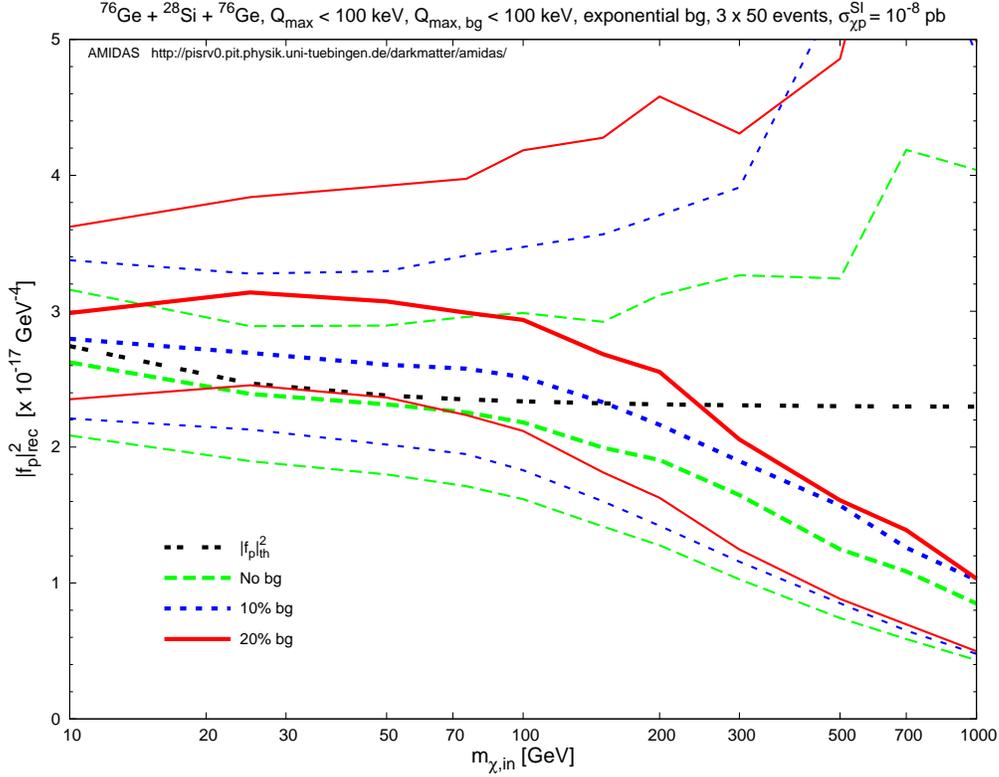} \\
\vspace{-0.5cm}
\end{center}
\caption{
 As in Fig.~\ref{fig:fp2-Ge-ex-000-100-050},
 except that
 the WIMP mass $\mchi$ in Eq.~(\ref{eqn:fp2})
 has been reconstructed by using two {\em other} data sets
 with a $\rmXA{Si}{28}$ target
 and a {\em second} $\rmXA{Ge}{76}$ target.
}
\label{fig:fp2-Ge-SiGe-ex-000-100-050}
\end{figure}

 In Fig.~\ref{fig:fp2-Ge-SiGe-ex-000-100-050}
 the WIMP mass $\mchi$ needed in expression (\ref{eqn:fp2})
 has been reconstructed by using two {\em other} data sets
 with a $\rmXA{Si}{28}$ target
 and a {\em second} $\rmXA{Ge}{76}$ target.
 As shown in Figs.~\ref{fig:mchi-SiGe-000-100-050}
 and discussed in Sec.~3.2,
 due to the contribution from residue background events,
 if the input WIMP mass is light (heavy),
 the reconstructed mass would be overestimated (underestimated).
 Hence,
 for input masses $\lsim$ ($\gsim$) 150 GeV,
 the SI WIMP--nucleon couplings
 reconstructed by using three independent data sets
 would be {\em larger} ({\em smaller}) than those
 reconstructed by using only one data set
 with an extra information about the WIMP mass
 (cf.~Fig.~\ref{fig:fp2-Ge-ex-000-100-050}).
 In addition,
 the statistical uncertainties
 on the reconstructed SI WIMP couplings
 would also be (much) {\em larger}.
 However,
 both Figs.~\ref{fig:fp2-Ge-ex-000-100-050}
 and \ref{fig:fp2-Ge-SiGe-ex-000-100-050}
 indicate that
 one could in principle estimate the SI WIMP--nucleon coupling
 up to a WIMP mass of $\sim$ 1 TeV
 by using one or three independent data sets
 with maximal 20\% background events
 (solid red).
 For a WIMP mass of 100 GeV
 and 20\% residue background events,
 the systematic deviation of
 the reconstructed SI WIMP coupling $|f_{\rm p}|$
 (with a reconstructed WIMP mass)
 would in principle be $\sim +13\%$
 with a statistical uncertainty
 of {$\sim^{+21\%}_{-14\%}$}
 ($\sim -3.3\%^{+18\%}_{-13\%}$
  for background--free data sets).

\begin{figure}[p!]
\begin{center}
\includegraphics[width=9.8cm]{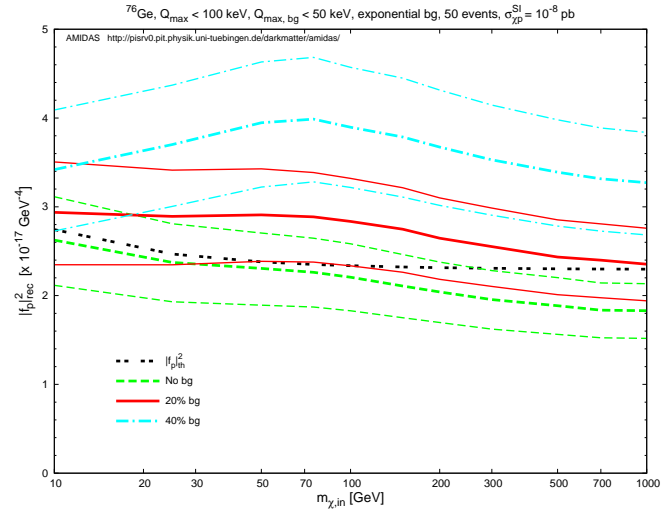}    \\
\vspace{0.75cm}
\includegraphics[width=9.8cm]{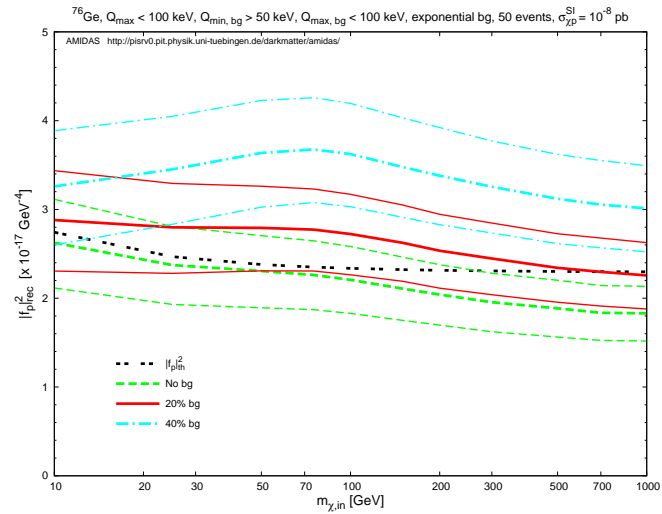}    \\
\vspace{0.75cm}
\includegraphics[width=9.8cm]{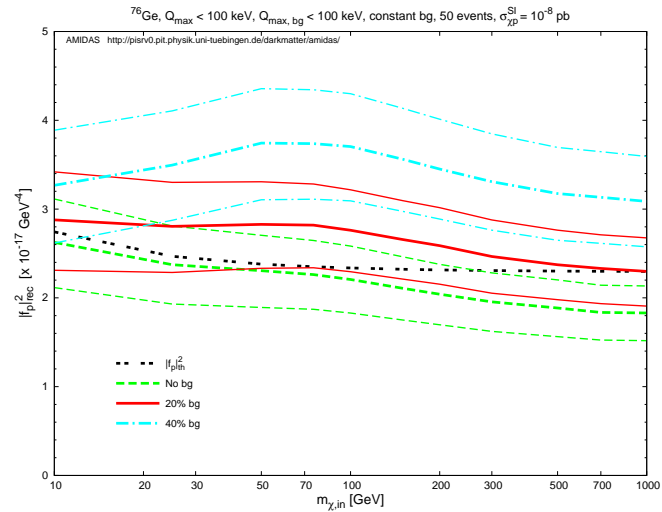} \\
\vspace{-0.25cm}
\end{center}
\caption{
 As in Fig.~\ref{fig:fp2-Ge-ex-000-100-050},
 except that the background spectra
 and/or windows are different:
 exponential  spectrum between  0 and  50 keV (top),
 exponential  spectrum between 50 and 100 keV (middle),
 and constant spectrum between  0 and 100 keV (bottom).
}
\label{fig:fp2-Ge-050}
\end{figure}
\begin{figure}[p!]
\begin{center}
\includegraphics[width=9.8cm]{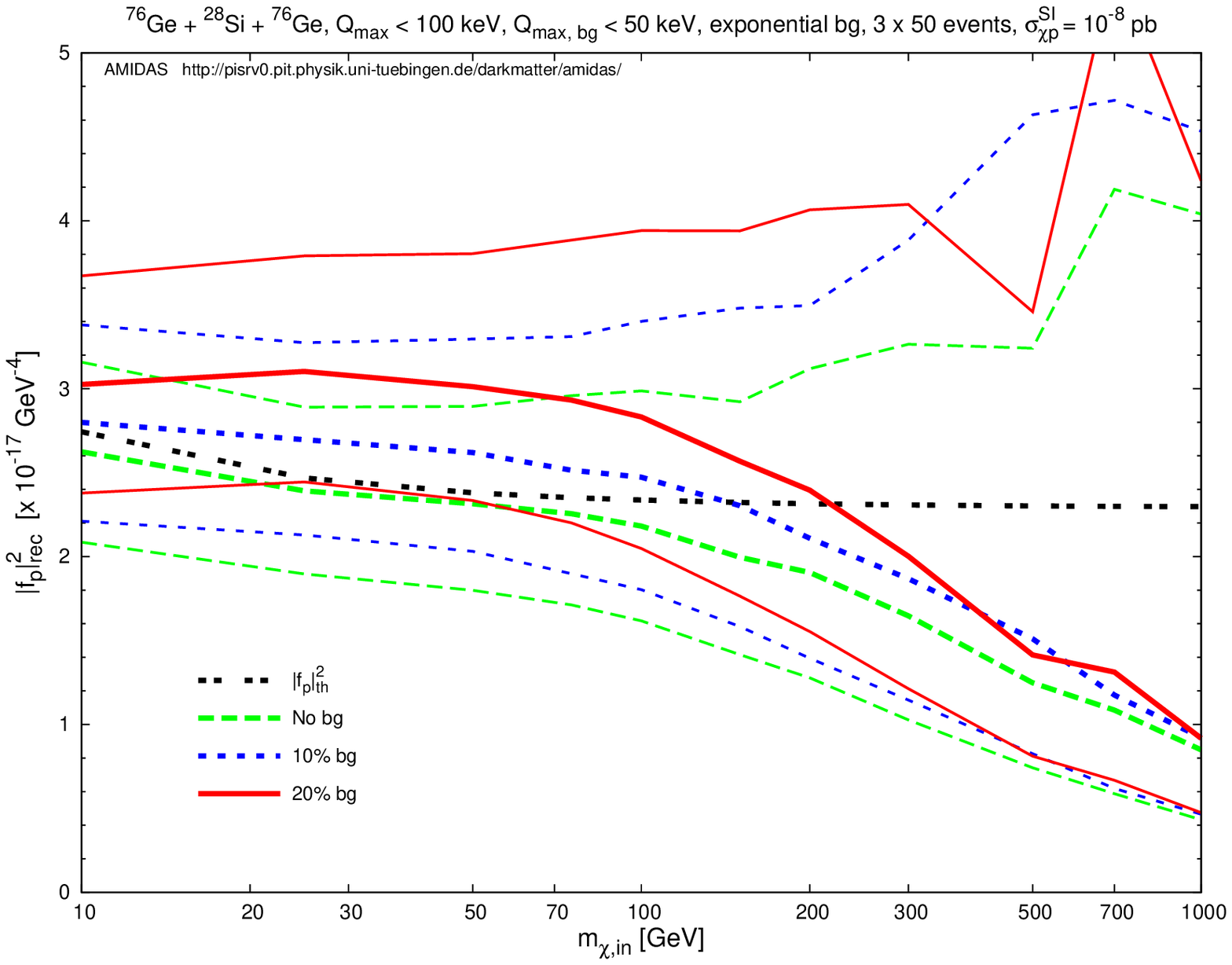}         \\
\vspace{0.75cm}
\includegraphics[width=9.8cm]{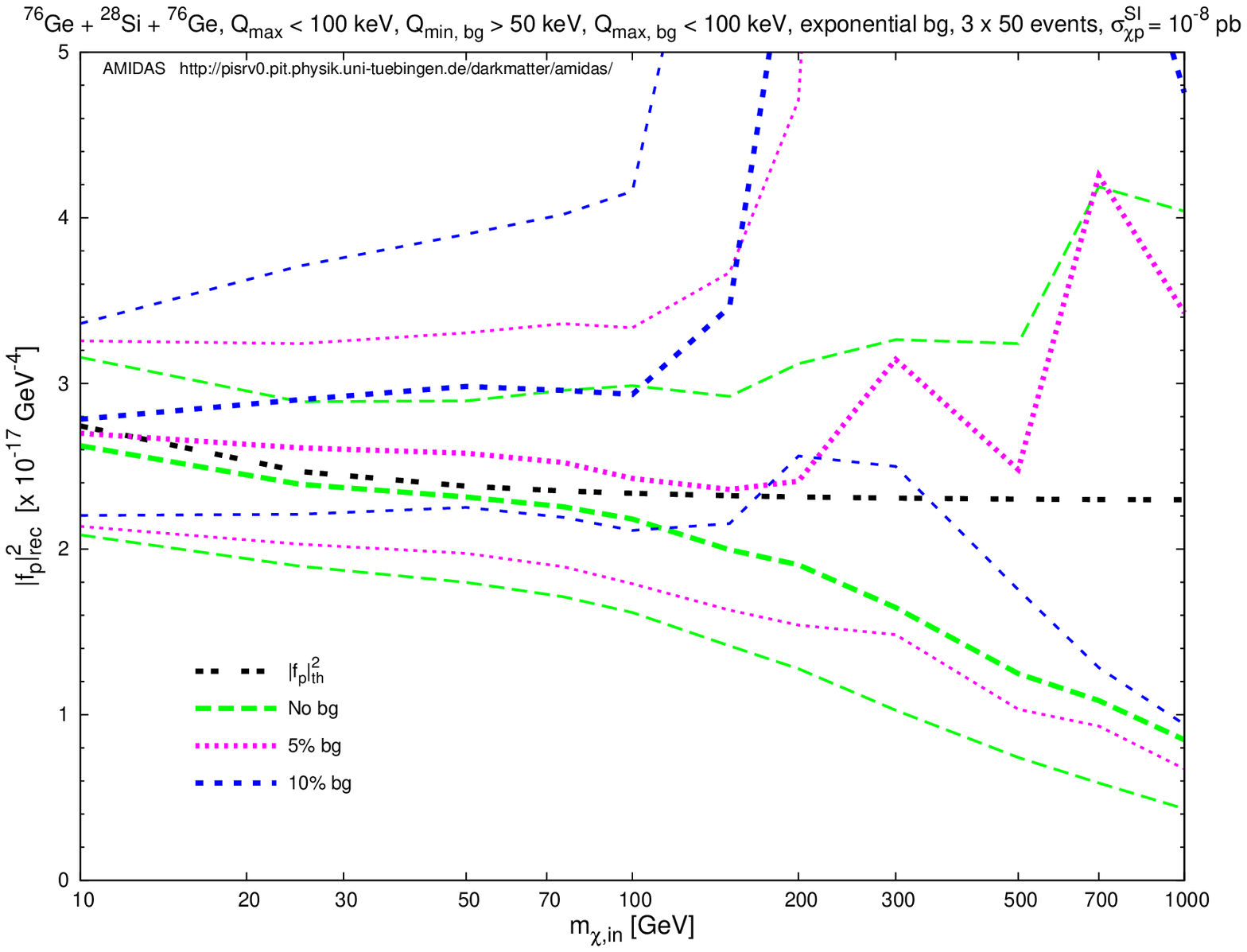}    \\
\vspace{0.75cm}
\includegraphics[width=9.8cm]{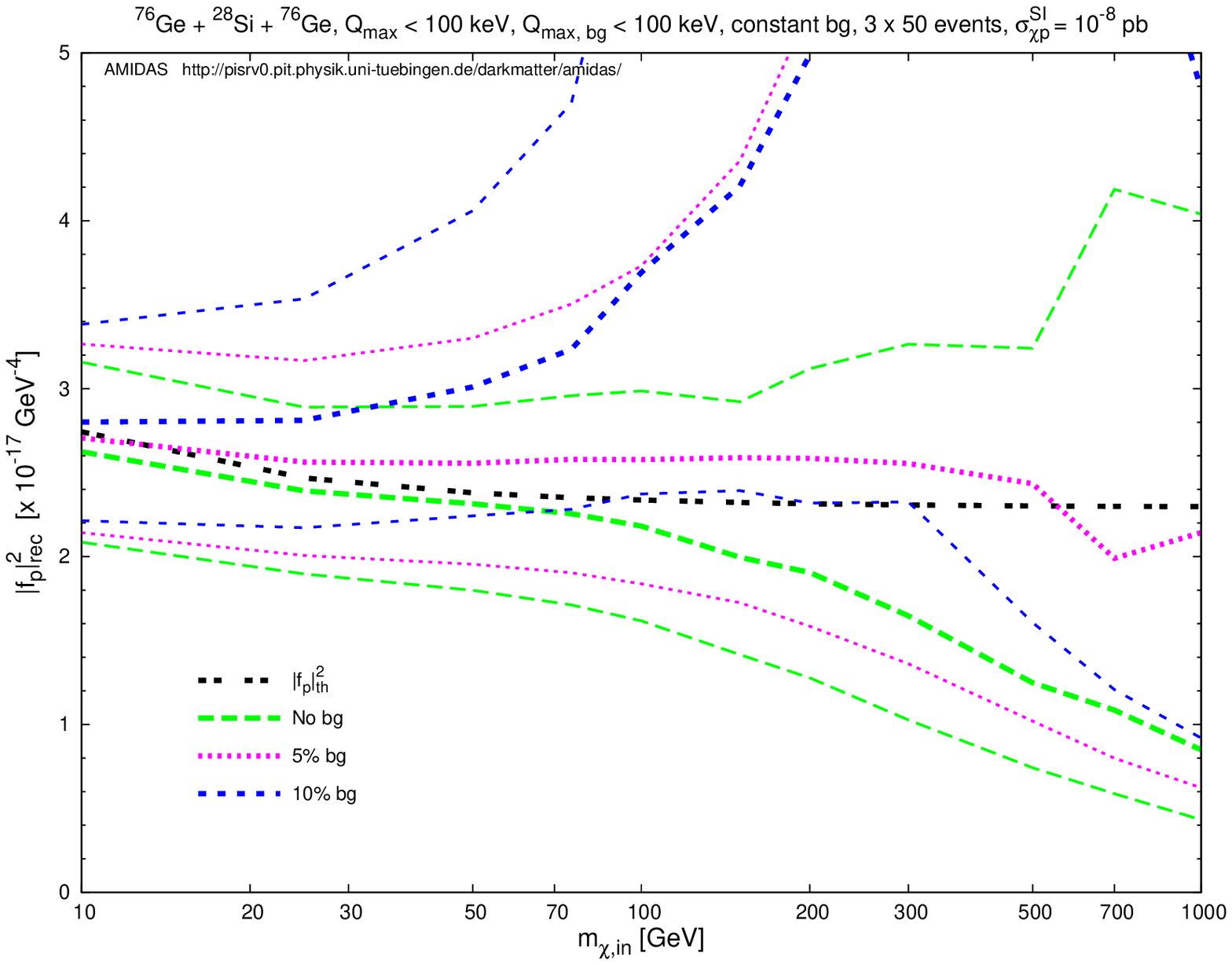} \\
\vspace{-0.25cm}
\end{center}
\caption{
 As in Figs.~\ref{fig:fp2-Ge-050},
 except that
 the WIMP mass $\mchi$ in Eq.~(\ref{eqn:fp2})
 has been reconstructed by using two {\em other} data sets
 with a $\rmXA{Si}{28}$ target
 and a {\em second} $\rmXA{Ge}{76}$ target.
 The background ratios shown in two lower frames
 are no background (dashed green),
  5\% (dotted magenta),
 and 10\% (long--dotted blue).
}
\label{fig:fp2-Ge-SiGe-050}
\end{figure}

 Furthermore,
 in order to check
 effects of different background discrimination ability
 in different energy ranges and
 the need of a prior knowledge about
 an (exact) form of the residue background spectrum,
 in Figs.~\ref{fig:fp2-Ge-050} and \ref{fig:fp2-Ge-SiGe-050}
 we consider
 two different background windows:
 between 0 and 50 keV and between 50 and 100 keV
 for the exponential background spectrum
 as well as
 the rather extrem constant spectrum
 given in Eq.~(\ref{eqn:dRdQ_bg_const})
 with a window between 0 and 100 keV.

 Firstly,
 from Figs.~\ref{fig:fp2-Ge-050},
 it can be seen clearly that,
 by using one data set
 with up to 20\% background events
 and a precisely known WIMP mass
 as an input information,
 the SI WIMP--nucleon coupling $|f_{\rm p}|$
 can in principle be estimated
 with a maximal $\sim +11\%$ systematic deviation
 (for an input WIMP mass of 100 GeV)
 from the theoretical value
 and a statistical uncertainty of $\sim \pm 9\%$.
 More importantly,
 all three cases show almost the same result.
 This indicates that,
 once the WIMP mass can be known (pretty) precisely,
 {\em not} the exact form of the residue background spectrum,
 but the {\em amount} in the analyzed data set
 could affect (significantly)
 the reconstructed SI WIMP--nucleon coupling.
 
 In contrast,
 results shown in Figs.~\ref{fig:fp2-Ge-SiGe-050}
 depend strongly on the reconstruction of the WIMP mass%
\footnote{
 Note that
 in our simulations shown here
 it was assumed that
 the spectra and windows of residue background events
 are the same for all three data sets.
 For practical use with different forms and windows
 of background events in different experiments,
 one can in principle follow the observations
 discussed in Sec.~3 and here.
}.
 As discussed in Ref.~\cite{DMDDbg-mchi} and Sec.~3.2,
 for cases with the exponential background spectrum
 and background windows
 in the whole experimental possible and low energy ranges,
 the reconstructed WIMP mass
 could be slightly overestimated (underestimated),
 once incident WIMPs are light (heavy).
 However,
 for the case with the exponential spectrum
 and windows in high energy ranges,
 or the case with the constant spectrum
 and windows of whole experimental possible energy ranges,
 the reconstructed WIMP mass
 could be (strongly) overestimated
 for all input WIMP masses.
 Here the effect of a (strongly) overestimated WIMP mass
 can be seen clearly here.
 Since for the case shown
 in Fig.~\ref{fig:fp2-Ge-SiGe-ex-000-100-050}
 our background spectra are exponential,
 only very few background events
 could be observed in the energy range
 between 50 and 100 keV.
 Hence,
 for the case with background windows
 only in the {\em low} energy ranges
 (top in Figs.~\ref{fig:fp2-Ge-SiGe-050}),
 not surprisingly,
 the result of the reconstructed SI WIMP coupling
 is almost the same as shown
 in Fig.~\ref{fig:fp2-Ge-SiGe-ex-000-100-050}.

 However,
 for the case with the exponential background spectrum
 and background windows in {\em high} energy ranges
 (middle in Figs.~\ref{fig:fp2-Ge-SiGe-050})
 or the case with the constant spectrum
 in whole experimental possible energy ranges
 (bottom),
 the results are almost the same:
 the larger the background ratio,
 the more strongly {\em overestimated} the SI WIMP coupling,
 in particular for heavy input WIMP masses.
 Nevertheless,
 by using (two or) three data sets
 with background ratios of $\lsim~10\%$,
 one could in principle reconstruct
 the SI WIMP--nucleon coupling
 (as well as the WIMP mass \cite{DMDDbg-mchi}) pretty well,
 without knowing the (exact) form of the background spectrum.

\begin{figure}[p!]
\begin{center}
\includegraphics[width=15cm]{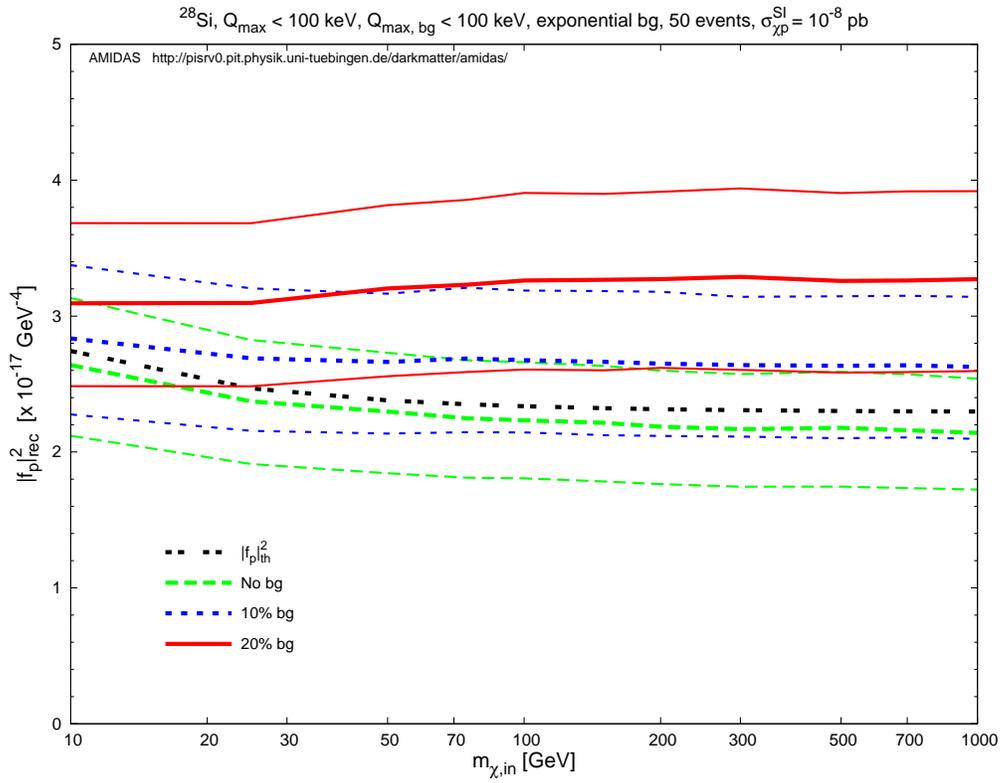} \\ \vspace{1cm}
\includegraphics[width=15cm]{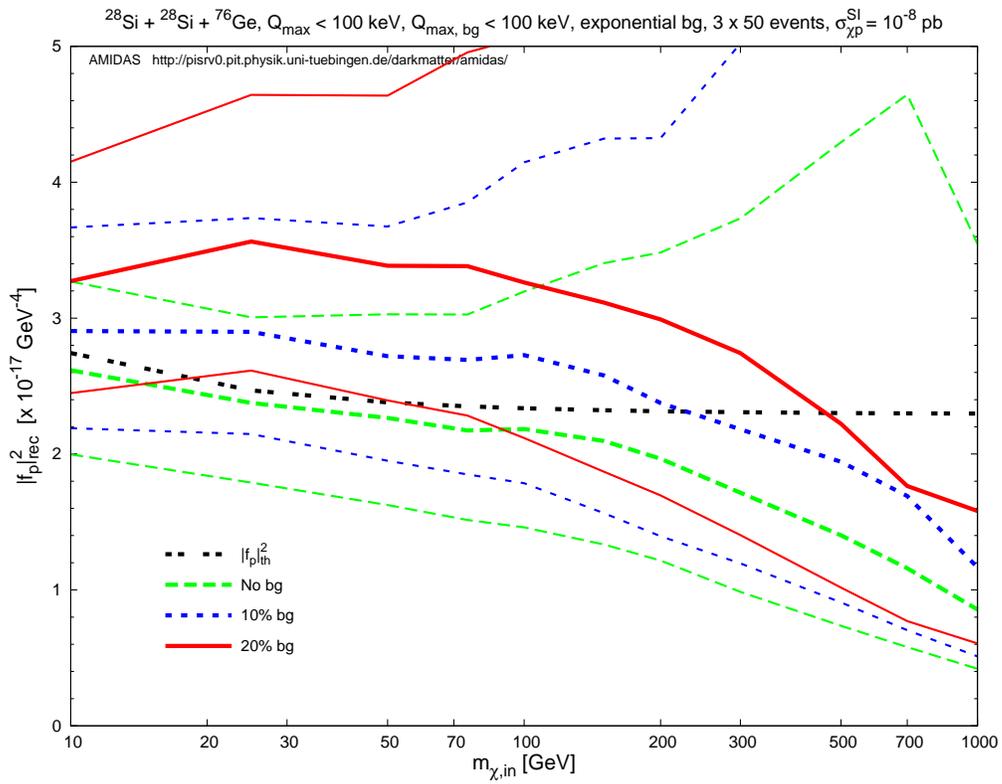} \\
\end{center}
\caption{
 As in Figs.~\ref{fig:fp2-Ge-ex-000-100-050}
 (upper) and
 \ref{fig:fp2-Ge-SiGe-ex-000-100-050}
 (lower),
 except that
 $\rmXA{Si}{28}$ has been used
 as the (first) target
 for reconstructing $f_{\rm p}|^2$.
 Remind that
 two data sets with the $\rmXA{Si}{28}$ target
 are {\em different}.
}
\label{fig:fp2-Si-SiGe-ex-000-100-050}
\end{figure}
\begin{figure}[t!]
\begin{center}
\includegraphics[width=12cm]{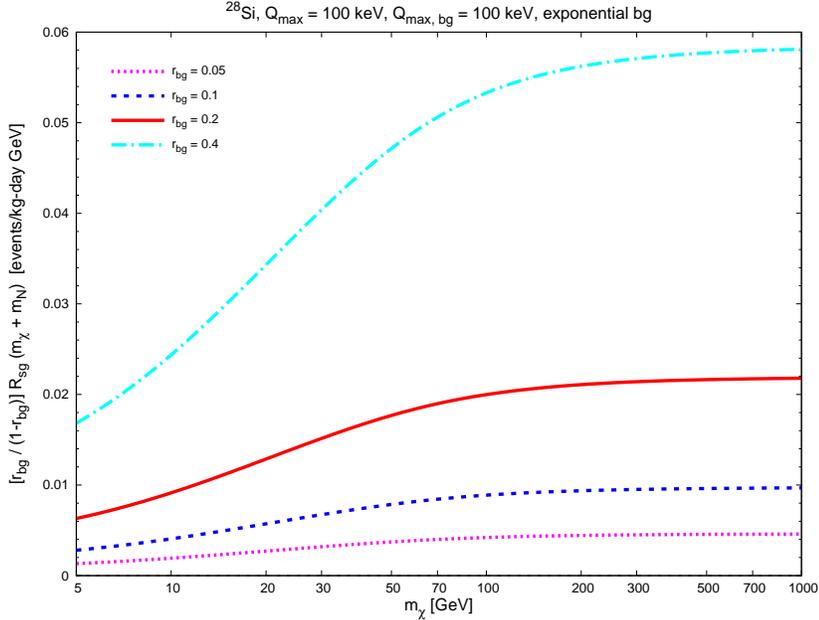} \\
\vspace{-0.5cm}
\end{center}
\caption{
 As in Fig.~\ref{fig:Rsg-mchi-Ge-000-100},
 except that
 a $\rmXA{Si}{28}$ target has been used.
}
\label{fig:Rsg-mchi-Si-000-100}
\end{figure}
\begin{figure}[p!]
\begin{center}
\vspace{-0.75cm}
\includegraphics[width=15cm]{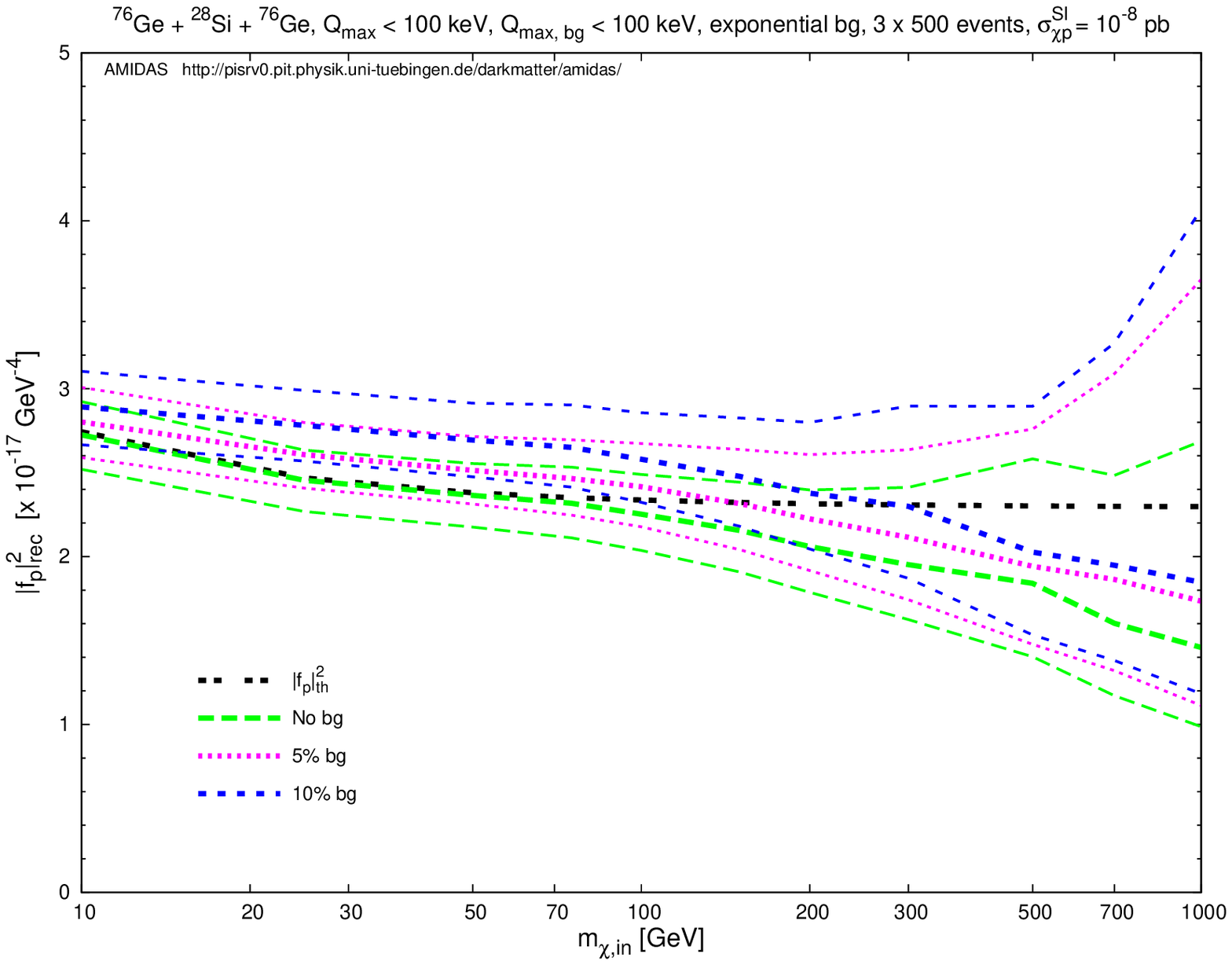} \\ \vspace{1cm}
\includegraphics[width=15cm]{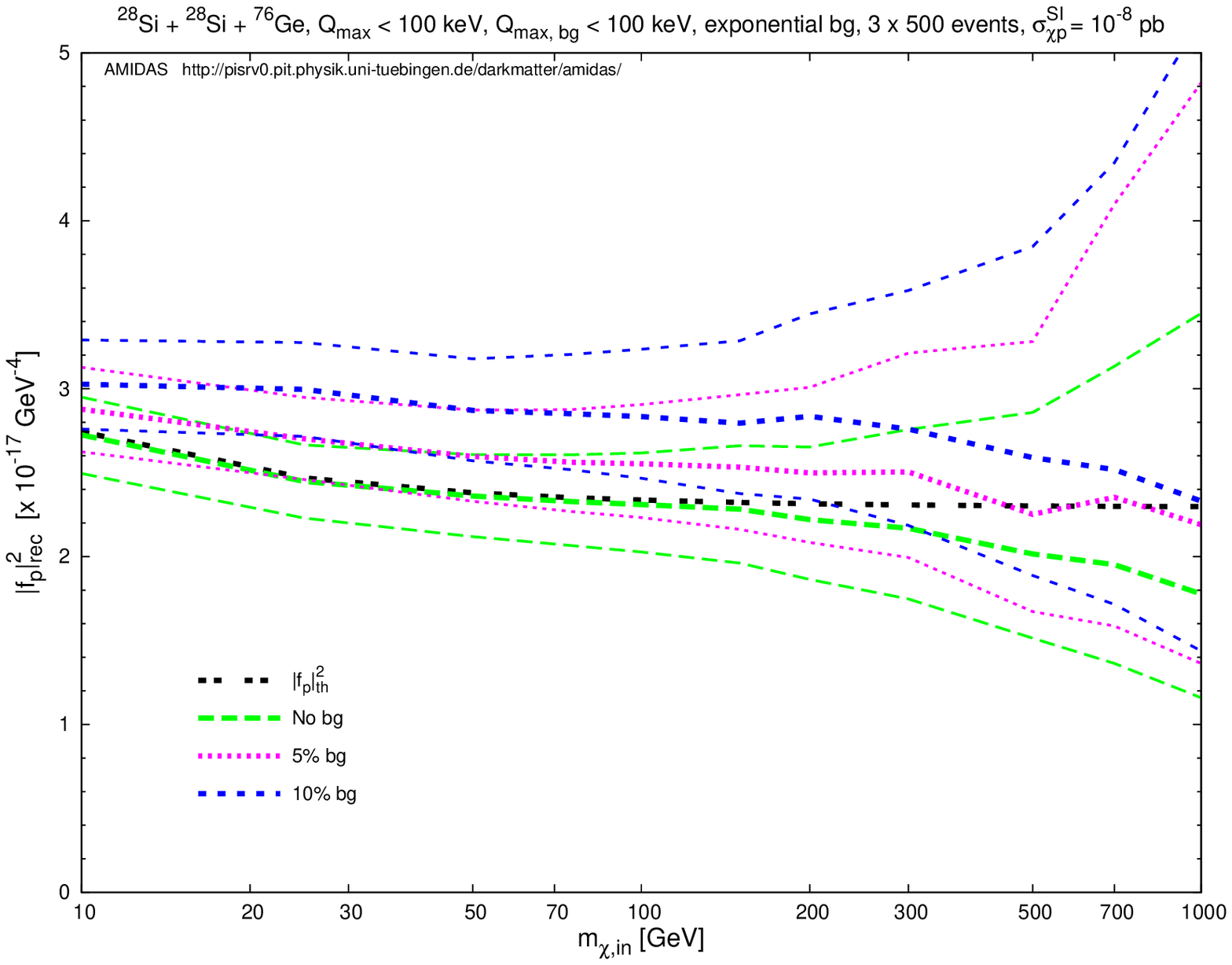} \\
\vspace{-0.25cm}
\end{center}
\caption{
 As in Fig.~\ref{fig:fp2-Ge-SiGe-ex-000-100-050}
 and the lower frame of
 Figs.~\ref{fig:fp2-Si-SiGe-ex-000-100-050},
 except that
 the expected number of total events
 in each data set
 has been risen by a factor of 10,
 i.e., 500 events on average,
 for both 
 the $\rmXA{Ge}{76}$ (upper) and
 the $\rmXA{Si}{28}$ (lower) target.
 The background ratios shown here
 are no background (dashed green),
  5\% (dotted magenta),
 and 10\% (long--dotted blue)
 background events in the analyzed data sets
 in the experimental energy ranges
 between 0 and 100 keV.
}
\label{fig:fp2-ex-000-100-500}
\end{figure}

 In Figs.~\ref{fig:fp2-Si-SiGe-ex-000-100-050}
 we consider a rather light target nucleus:
 $\rmXA{Si}{28}$.
 The WIMP mass $\mchi$
 has been assumed to be known precisely (upper)
 or reconstructed from other two data sets (lower).
 Only the exponential form for residue background spectrum
 has been considered here.
 Remind that,
 as found in Refs.~\cite{DMDDfp2-IDM2008, DMDDfp2},
 with a light target nucleus,
 e.g., Si or Ar,
 the statistical uncertainty
 on the reconstructed SI WIMP--nucleon coupling
 is {\em larger} than that with a heavy nucleus,
 e.g., Ge or Xe.
 Consequently,
 for both cases
 (with a precisely known
  or a reconstructed WIMP mass),
 the reconstructed SI WIMP couplings
 as well as the 1$\sigma$ statistical uncertainties
 shown in Figs.~\ref{fig:fp2-Si-SiGe-ex-000-100-050}
 are larger than those with $\rmXA{Ge}{76}$
 shown in Figs.~\ref{fig:fp2-Ge-ex-000-100-050}
 and \ref{fig:fp2-Ge-SiGe-ex-000-100-050}.

 On the other hand,
 the systematic deviations of
 the {\em (under)estimated} SI WIMP coupling
 for {\em heavy} input WIMP masses
 ($\mchi~\gsim~300$ GeV)
 are {\em smaller} for light nuclei than
 those for heavy ones
 \cite{DMDDfp2-IDM2008, DMDDfp2}.
 In addition,
 as shown in Fig.~\ref{fig:Rsg-mchi-Si-000-100},
 the background contribution
 (the second term on the right--hand side of Eq.~(\ref{eqn:rmin_I0}))
 increases pretty quickly
 with an increased WIMP mass.
 Hence,
 for input WIMP masses \mbox{$\mchi~\gsim~100$ GeV},
 the reconstructed SI WIMP--nucleon couplings
 with a $\rmXA{Si}{28}$ target nucleus
 would be {\em more strongly overestimated} than
 those reconstructed with a $\rmXA{Ge}{76}$ target.
 Nevertheless,
 for both cases
 (with a precisely known
  or a reconstructed WIMP mass),
 with $\sim$ 10\% -- 20\% background events in our data sets,
 the 1$\sigma$ statistical uncertainty bands
 could in principle still cover
 the theoretical value of $|f_{\rm p}|$.
 For an input WIMP mass of 100 GeV,
 by using three data sets with 10\% background events,
 the systematic deviation of and the statistical uncertainty on
 the reconstructed SI WIMP--nucleon coupling $|f_{\rm p}|$ is
 $\sim +8\%^{+26\%}_{-17\%}$
 ($\sim -3\%^{+23\%}_{-17\%}$
  for background--free data sets).

 Finally,
 considering the progress of
 detection and background discrimination techniques
 of the next--generation ton--scale detectors,
 in Figs.~\ref{fig:fp2-ex-000-100-500}
 we rise the expected number of total events
 in each data set by a factor of 10,
 i.e., 500 events on average,
 for both 
 $\rmXA{Ge}{76}$ (upper) and
 $\rmXA{Si}{28}$ (lower) targets.
 Here I show only the results
 with the reconstructed WIMP mass.
 Since the statistical uncertainties
 shrink now by a factor of $\gsim~3$,
 the maximal acceptable background ratio becomes $\sim$ 5\%
 (i.e., $\sim$ 25 residue background events in each data set).
 For an input WIMP mass of 100 GeV,
 the systematic deviation would then be $\sim +2\%$ (with Ge)
 and $\sim +5\%$ (with Si)
 with a statistical uncertainty of
 $\sim 5\%$ (with Ge)
 and $\sim 7\%$ (with Si).
\section{Summary and conclusions}
 In this paper
 I reexamine the model--independent data analysis method
 introduced in Refs.~\cite{DMDDfp2-IDM2008, DMDDfp2}
 for the estimation of
 the spin--independent scalar coupling
 of Weakly Interacting Massive Particles
 on nucleons
 from data (measured recoil energies) of
 direct Dark Matter detection experiments directly
 by taking into account a fraction of residue background events,
 which pass all discrimination criteria and
 then mix with other real WIMP--induced events
 in the analyzed data sets.
 This method requires {\em neither} prior knowledge
 about the WIMP scattering spectrum
 {\em nor} about different possible background spectra;
 the needed information is the recoil energies
 recorded in direct detection experiments,
 an unique assumption about the local WIMP density,
 and (occasionally) the mass of incident WIMPs.

 For the mass of incident WIMPs
 required in this data analysis,
 we considered two cases:
 known precisely with an overall uncertainty
 (of 5\% of the input WIMP mass in our simulations)
 from other (e.g., collider) experiments
 as well as reconstructed
 by using other direct detection experiments.
 Our simulations show that,
 assuming an exponential form
 for the residue background spectrum,
 with $\sim$ 50 total events in each data set,
 for both cases
 the maximal acceptable background ratio
 is $\sim$ 20\%
 (i.e., $\sim$ 10 background events).
 For a WIMP mass of 100 GeV
 and 20\% residue background events,
 the systematic deviation of
 the reconstructed SI WIMP coupling $|f_{\rm p}|$
 (with a reconstructed WIMP mass)
 would in principle be $\sim +13\%$
 with a statistical uncertainty
 of {$\sim^{+21\%}_{-14\%}$}
 ($\sim -3.3\%^{+18\%}_{-13\%}$
  for background--free data sets).

 Furthermore,
 in order to check
 effects of different background discrimination ability
 in different energy ranges and
 the need of a prior knowledge about
 an (exact) form of the residue background spectrum,
 we considered two different background windows:
 between 0 and 50 keV and between 50 and 100 keV
 for the exponential background spectrum
 as well as the rather extrem constant spectrum
 with a window between 0 and 100 keV.
 It has been found that,
 with a precisely known WIMP mass,
 all three cases show almost the same result
 as that for the exponential background spectrum
 with a window of the whole experimental possible energy range.
 This indicates that,
 once the WIMP mass can be known (pretty) precisely,
 {\em not} the exact form of the residue background spectrum,
 but their {\em amount} in the analyzed data set
 could affect (significantly)
 the reconstructed SI WIMP--nucleon coupling.

 On the other hand,
 we considered a rather light target nucleus:
 $\rmXA{Si}{28}$
 with both a precisely known
 and a reconstructed WIMP mass.
 For both cases
 the reconstructed SI WIMP couplings
 would be more strongly overestimated
 and the 1$\sigma$ statistical uncertainties
 would also be larger than those with $\rmXA{Ge}{76}$.
 Nevertheless,
 with $\sim$ 10\% -- 20\% background events in our data sets,
 the 1$\sigma$ statistical uncertainty bands
 could in principle still cover the theoretical value of $|f_{\rm p}|$.
 For an input WIMP mass of 100 GeV,
 by using three data sets with 10\% background events,
 the systematic deviation of and the statistical uncertainty on
 the reconstructed SI WIMP--nucleon coupling $|f_{\rm p}|$ is
 $\sim +8\%^{+26\%}_{-17\%}$
 ($\sim -3\%^{+23\%}_{-17\%}$
  for background--free data sets).

 Finally,
 for rather next--generation ton--scale detectors,
 we considered the use of data sets of ${\cal O}(500)$ events
 for both $\rmXA{Ge}{76}$ and $\rmXA{Si}{28}$ targets.
 Our results show that,
 with a maximal background ratio of 5\%
 (i.e., $\sim$ 25 total events in each data set),
 one could in principle still reconstructed
 the SI WIMP--nucleon coupling pretty well:
 for an input WIMP mass of 100 GeV,
 the systematic deviation would be $\sim +2\%$ (with Ge)
 and $\sim +5\%$ (with Si)
 with a statistical uncertainty of
 $\sim 5\%$ (with Ge)
 and $\sim 7\%$ (with Si).

 In summary,
 as the third part of
 the study of the effects of residue background events
 in direct Dark Matter detection experiments,
 we considered the estimation of
 the SI WIMP--nucleon coupling.
 Our results show that,
 with currently running and projected experiments
 using detectors with $10^{-9}$ to $10^{-11}$ pb sensitivities
 \cite{Baudis07a, Aprile09a, Gascon09, Drees10}
 and $< 10^{-6}$ background rejection ability
 \cite{CRESST-bg, EDELWEISS-bg, Lang09b, Ahmed09b},
 once one or more experiments with different target nuclei
 could accumulate a few tens events
 (in one experiment),
 we could in principle already estimate
 the SI coupling of Dark Matter particles on ordinary matter
 with a reasonable precession,
 or at least give an upper bound on that,
 even though there could be some background events
 mixed in our data sets for the analysis
 and the reconstructed value would thus be overestimated.
 Moreover,
 although two forms for background spectrum and
 three windows for residue background events
 considered in this work are rather naive;
 one should be able to extend our observations/discussions
 to predict the effects of possible background events
 in their own experiment.
 Hopefully,
 this will encourage our experimental colleagues
 to present their (future) results
 as the ``most possible area(s)''
 in the parameter space of different extensions of
 the Standard Model of particle physics
 and in turn to offer stringenter information
 for identifying (WIMP) Dark Matter particles at colliders
 as well as
 for predicting spectra
 in indirect Dark Matter detection experiments.
\subsubsection*{Acknowledgments}
 The author would like to thank
 the Physikalisches Institut der Universit\"at T\"ubingen
 for the technical support of the computational work
 demonstrated in this article.
 This work
 was partially supported by
 the National Science Council of R.O.C.~%
 under contract no.~NSC-99-2811-M-006-031
 as well as by
 the LHC Physics Focus Group,
% the Focus Group on Cosmology and Particle Astrophysics,
 National Center of Theoretical Sciences, R.O.C..
\appendix
\setcounter{equation}{0}
\setcounter{figure}{0}
\renewcommand{\theequation}{A\arabic{equation}}
\renewcommand{\thefigure}{A\arabic{figure}}
%
%
% Appendix A
\section{Formulae needed in Sec.~2}
 Here I list all formulae needed
 for the model--independent data analysis procedure
 used in Sec.~2.
 Detailed derivations and discussions
 can be found in Refs.~\cite{DMDDf1v, DMDDmchi}.
\subsection{Estimating \boldmath$r(\Qmin)$ and $I_n(\Qmin, \Qmax)$}
 First,
 consider experimental data described by
\beq
     {\T Q_n - \frac{b_n}{2}}
 \le \Qni
 \le {\T Q_n + \frac{b_n}{2}}
\~,
     ~~~~~~~~~~~~ %12
     i
 =   1,~2,~\cdots,~N_n,~
     n
 =   1,~2,~\cdots,~B.
\label{eqn:Qni}
\eeq
 Here the total energy range between $\Qmin$ and $\Qmax$
 has been divided into $B$ bins
 with central points $Q_n$ and widths $b_n$.
 In each bin,
 $N_n$ events will be recorded.
 Since the recoil spectrum $dR / dQ$ is expected
 to be approximately exponential, %\cite{DMDDf1v},
 the following ansatz for the {\em measured} recoil spectrum
 ({\em before} normalized by the experimental exposure $\calE$)
 in the $n$th bin has been introduced \cite{DMDDf1v}:
\beq
        \adRdQ_{{\rm expt}, \~ n}
 \equiv \adRdQ_{{\rm expt}, \~ Q \simeq Q_n}
 \equiv \rn  \~ e^{k_n (Q - Q_{s, n})}
\~.
\label{eqn:dRdQn}
\eeq
 Here $r_n$ is the standard estimator
 for $(dR / dQ)_{\rm expt}$ at $Q = Q_n$:
\beq
   r_n
 = \frac{N_n}{b_n}
\~,
\label{eqn:rn}
\eeq
 $k_n$ is the logarithmic slope of
 the recoil spectrum in the $n$th $Q-$bin,
 which can be computed numerically
 from the average value of the measured recoil energies
 in this bin:
\beq
   \bQn
 = \afrac{b_n}{2} \coth\afrac{k_n b_n}{2}-\frac{1}{k_n}
\~,
\label{eqn:bQn}
\eeq
 where
\beq
        \bQxn{\lambda}
 \equiv \frac{1}{N_n} \sumiNn \abrac{\Qni - Q_n}^{\lambda}
\~.
\label{eqn:bQn_lambda}
\eeq
 The error on the logarithmic slope $k_n$
 can be estimated from Eq.~(\ref{eqn:bQn}) directly as
\beq
   \sigma^2(k_n)
 = k_n^4
   \cbrac{  1
          - \bfrac{k_n b_n / 2}{\sinh (k_n b_n / 2)}^2}^{-2}
            \sigma^2\abrac{\bQn}
\~,
\label{eqn:sigma_kn}
\eeq
 with
\beq
   \sigma^2\abrac{\bQn}
 = \frac{1}{N_n - 1} \bbigg{\bQQn - \bQn^2}
\~.
\label{eqn:sigma_bQn}
\eeq
 $Q_{s, n}$ in the ansatz (\ref{eqn:dRdQn})
 is the shifted point at which
 the leading systematic error due to the ansatz
 is minimal \cite{DMDDf1v},
\beq
   Q_{s, n}
 = Q_n + \frac{1}{k_n} \ln\bfrac{\sinh(k_n b_n/2)}{k_n b_n/2}
\~.
\label{eqn:Qsn}
\eeq
 Note that $Q_{s, n}$ differs from
 the central point of the $n$th bin, $Q_n$.
 From the ansatz (\ref{eqn:dRdQn}),
 the counting rate at $Q = \Qmin$ can be calculated by
\beq
   r(\Qmin)
 = r_1 e^{k_1 (\Qmin - Q_{s, 1})}
\~,
\label{eqn:rmin_eq}
\eeq
 and its statistical error can be expressed as
\beq
   \sigma^2(r(\Qmin))
 = r^2(\Qmin) 
   \cbrac{  \frac{1}{N_1}
          + \bbrac{  \frac{1}{k_1}
                   - \afrac{b_1}{2} 
                     \abrac{  1
                            + \coth\afrac{b_1 k_1}{2}}}^2
            \sigma^2(k_1)}
\~,
\label{eqn:sigma_rmin}
\eeq
 since
\beq
   \sigma^2(r_n)
 = \frac{N_n}{b_n^2}
\~.
\label{eqn:sigma_rn}
\eeq
 Finally,
 since all $I_n$ are determined from the same data,
 they are correlated with
\beq
   {\rm cov}(I_n, I_m)
 = \sum_{a = 1}^{N_{\rm tot}} \frac{Q_a^{(n+m-2)/2}}{F^4(Q_a)}
\~,
\label{eqn:cov_In}
\eeq
 where the sum runs over all events
 with recoil energy between $\Qmin$ and $\Qmax$. 
 And the correlation between the errors on $r(\Qmin)$,
 which is calculated entirely
 from the events in the first bin,
 and on $I_n$ is given by
\beqn
 \conti {\rm cov}(r(\Qmin), I_n)
        \non\\
 \=     r(\Qmin) \~ I_n(\Qmin, \Qmin + b_1)
        \non\\
 \conti ~~~~ \times %4
        \cleft{  \frac{1}{N_1} 
               + \bbrac{  \frac{1}{k_1}
                        - \afrac{b_1}{2} \abrac{1 + \coth\afrac{b_1 k_1}{2}}}}
        \non\\
 \conti ~~~~~~~~~~~~~~ \times %14
        \cright{ \bbrac{  \frac{I_{n+2}(\Qmin, \Qmin + b_1)}
                               {I_{n  }(\Qmin, \Qmin + b_1)}
                        - Q_1
                        + \frac{1}{k_1}
                        - \afrac{b_1}{2} \coth\afrac{b_1 k_1} {2}}
        \sigma^2(k_1)}
\~;
\label{eqn:cov_rmin_In}
\eeqn
 note that
 the sums $I_i$ here only count in the first bin,
 which ends at $Q = \Qmin + b_1$.

 On the other hand,
 with a functional form of the recoil spectrum
 (e.g., fitted to experimental data),
 $(dR / dQ)_{\rm expt}$,
 one can use the following integral forms
 to replace the summations given above.
 Firstly,
 the average $Q-$value in the $n$th bin
 defined in Eq.~(\ref{eqn:bQn_lambda})
 can be calculated by
\beq
   \bQxn{\lambda}
 = \frac{1}{N_n} \intQnbn \abrac{Q - Q_n}^{\lambda} \adRdQ_{\rm expt} dQ
\~.
\label{eqn:bQn_lambda_int}
\eeq
 For $I_n(\Qmin, \Qmax)$ given in Eq.~(\ref{eqn:In_sum}),
 we have
\beq
   I_n(\Qmin, \Qmax)
 = \int_{\Qmin}^{\Qmax} \frac{Q^{(n-1)/2}}{F^2(Q)} \adRdQ_{\rm expt} dQ
\~,
\label{eqn:In_int}
\eeq 
 and similarly for the covariance matrix for $I_n$
 in Eq.~(\ref{eqn:cov_In}),
\beq
   {\rm cov}(I_n, I_m)
 = \int_{\Qmin}^{\Qmax} \frac{Q^{(n+m-2)/2}}{F^4(Q)} \adRdQ_{\rm expt} dQ
\~.
\label{eqn:cov_In_int}
\eeq 
 Remind that
 $(dR / dQ)_{\rm expt}$ is the {\em measured} recoil spectrum
 {\em before} normalized by the exposure.
 Finally,
 $I_i(\Qmin, \Qmin + b_1)$ needed in Eq.~(\ref{eqn:cov_rmin_In})
 can be calculated by
\beq
   I_n(\Qmin, \Qmin + b_1)
 = \int_{\Qmin}^{\Qmin + b_1}
   \frac{Q^{(n-1)/2}}{F^2(Q)} \bbigg{r_1 \~ e^{k_1 (Q - Q_{s, 1})}} dQ
\~.
\label{eqn:In_1_int}
\eeq 
 Note that,
 firstly,
 $r(\Qmin)$ and $I_n(\Qmin, \Qmin + b_1)$ should be
 estimated by Eqs.~(\ref{eqn:rmin_eq}) and (\ref{eqn:In_1_int})
 with $r_1$, $k_1$ and $Q_{s, 1}$
 estimated by Eqs.~(\ref{eqn:rn}), (\ref{eqn:bQn}), and (\ref{eqn:Qsn})
 in order to use the other formulae for estimating
 the (correlations between the) statistical errors
 without any modification.
 Secondly,
 $r(\Qmin)$ and $I_n(\Qmin, \Qmax)$ estimated
 from a scattering spectrum fitted to experimental data
 are usually not model--independent any more.
\subsection{Determining the WIMP mass \boldmath$\mchi$}
 By requiring that
 the values of a given moment of $f_1(v)$
 estimated by Eq.~(\ref{eqn:moments})
 from two experiments
 with different target nuclei, $X$ and $Y$, agree,
 $\mchi$ appearing in the prefactor $\alpha^n$
 on the right--hand side of Eq.~(\ref{eqn:moments})
 can be solved analytically as
 \cite{DMDDmchi-SUSY07, DMDDmchi}:
\beq
   \left. \mchi \right|_{\Expv{v^n}}
 = \frac{\sqrt{\mX \mY} - \mX (\calR_{n, X} / \calR_{n, Y})}
        {\calR_{n, X} / \calR_{n, Y} - \sqrt{\mX / \mY}}
\~,
\label{eqn:mchi_Rn}
\eeq
 with $\calR_{n, (X, Y)}$ defined by
\beqn
        \calR_{n, X}
 \equiv \bfrac{2 \QminX^{(n + 1) / 2} r_X(\QminX) / \FQminX + (n + 1) \InX}
              {2 \QminX^{     1  / 2} r_X(\QminX) / \FQminX +         \IzX}^{1/n}
\~,
\label{eqn:RnX_min}
\eeqn
 and $\calR_{n, Y}$ can be defined analogously.
 Here $n \ne 0$,
 $m_{(X, Y)}$ and $F_{(X, Y)}(Q)$
 are the masses and the form factors of the nucleus $X$ and $Y$,
 respectively,
 and $r_{(X, Y)}(Q_{{\rm min}, (X, Y)})$
 refer to the counting rates for the target $X$ and $Y$
 at the respective lowest recoil energies included in the analysis.
 Note that
 the general expression (\ref{eqn:mchi_Rn}) can be used
 either for spin--independent or
 for spin--dependent scattering,
 one only needs to choose different form factors
 under different assumptions;
 the form factors needed for estimating $I_{n, (X, Y)}$
 by Eq.~(\ref{eqn:In_sum}) or (\ref{eqn:In_int})
 are thus also different.

 By using the standard Gaussian error propagation,
 a lengthy expression for the statistical uncertainty on
 $\left. \mchi \right|_{\Expv{v^n}}$
 given in Eq.~(\ref{eqn:mchi_Rn})
 can be obtained as
\beqn
        \left. \sigma(\mchi) \right|_{\Expv{v^n}}
 \=     \frac{\sqrt{\mX / \mY} \vbrac{\mX - \mY} \abrac{\calR_{n, X} / \calR_{n, Y}} }
             {\abrac{\calR_{n, X} / \calR_{n, Y} - \sqrt{\mX / \mY}}^2}
        \non\\
 \conti ~~~~ \times
        \bbrac{  \frac{1}{\calR_{n, X}^2}
                 \sum_{i, j = 1}^3
                 \aPp{\calR_{n, X}}{c_{i, X}} \aPp{\calR_{n, X}}{c_{j, X}}
                 {\rm cov}(c_{i, X}, c_{j, X})
               + (X \lto Y)}^{1 / 2}
\!.
\label{eqn:sigma_mchi_Rn}
\eeqn
 Here a short--hand notation for the six quantities
 on which the estimate of $\mchi$ depends
 has been introduced:
\beq
   c_{1, X}
 = I_{n, X}
\~,
   ~~~~~~~~~~~~ %12
   c_{2, X}
 = I_{0, X}
\~,
   ~~~~~~~~~~~~ %12
   c_{3, X}
 = r_X(\QminX)
\~;
\label{eqn:ciX}
\eeq
 and similarly for the $c_{i, Y}$.
 Estimators for ${\rm cov}(c_i, c_j)$ have been given
 in Eqs.~(\ref{eqn:cov_In}) and (\ref{eqn:cov_rmin_In}).
 Explicit expressions for the derivatives of $\calR_{n, X}$
 with respect to $c_{i, X}$ are:
\cheqnXa{A}
\beq
   \Pp{\calR_{n, X}}{\InX}
 = \frac{n + 1}{n}
   \bfrac{\FQminX}{2 \QminX^{(n + 1) / 2} r_X(\QminX) + (n + 1) \InX \FQminX}
   \calR_{n, X}
\~,
\label{eqn:dRnX_dInX}
\eeq
\cheqnXb{A}
\beq
   \Pp{\calR_{n, X}}{\IzX}
 =-\frac{1}{n}
   \bfrac{\FQminX}{2 \QminX^{1 / 2} r_X(\QminX) + \IzX \FQminX}
   \calR_{n, X}
\~,
\label{eqn:dRnX_dIzX}
\eeq
 and
\cheqnXc{A}
\beqn
        \Pp{\calR_{n, X}}{r_X(\QminX)}
 \=     \frac{2}{n}
        \bfrac{  \QminX^{(n + 1) / 2} \IzX        - (n + 1) \QminX^{1 / 2} \InX}
              {2 \QminX^{(n + 1) / 2} r_X(\QminX) + (n + 1) \InX \FQminX}
        \non\\
 \conti ~~~~~~~~~~~~~~~~ \times %16
        \bfrac{\FQminX}{2 \QminX^{1 / 2} r_X(\QminX) + \IzX \FQminX}
        \calR_{n, X}
\~;
\label{eqn:dRnX_drminX}
\eeqn
\cheqnX{A}%
 explicit expressions for the derivatives of $\calR_{n, Y}$
 with respect to $c_{i, Y}$ can be given analogously.
 Note that,
 firstly,
 factors $\calR_{n, (X, Y)}$ appear in all these expressions,
 which can practically be cancelled by the prefactors
 in the bracket in Eq.~(\ref{eqn:sigma_mchi_Rn}).
 Secondly,
 all the $I_{0, (X, Y)}$ and $I_{n, (X, Y)}$ should be understood
 to be computed according to
 Eq.~(\ref{eqn:In_sum}) or (\ref{eqn:In_int})
 with integration limits $\Qmin$ and $\Qmax$
 specific for that target.

 On the other hand,
 since $|f_{\rm p}|^2$ in Eq.~(\ref{eqn:fp2})
 is identical for different targets,
 it leads to a second expression for determining $\mchi$
 \cite{DMDDmchi}:
\beq
   \left. \mchi \right|_\sigma
 = \frac{\abrac{\mX / \mY}^{5 / 2} \mY - \mX (\calR_{\sigma, X} / \calR_{\sigma, Y})}
        {\calR_{\sigma, X} / \calR_{\sigma, Y} - \abrac{\mX / \mY}^{5 / 2}}
\~.
\label{eqn:mchi_Rsigma}
\eeq
 Here $m_{(X, Y)} \propto A_{(X, Y)}$ has been assumed,
 and $\calR_{\sigma, (X, Y)}$ are defined by
\beq
        \calR_{\sigma, X}
 \equiv \frac{1}{\calE_X}
        \bbrac{\frac{2 \QminX^{1 / 2} r_X(\QminX)}{\FQminX} + \IzX}
\~,
\label{eqn:RsigmaX_min}
\eeq
 and similarly for $\calR_{\sigma, Y}$;
 $\calE_{(X, Y)}$ here are the experimental exposures
 with the target $X$ and $Y$.
 Similar to the analogy between
 Eqs.~(\ref{eqn:mchi_Rn}) and (\ref{eqn:mchi_Rsigma}),
 the statistical uncertainty on $\left. \mchi \right|_\sigma$
 given in Eq.~(\ref{eqn:mchi_Rsigma})
 can be expressed as
\beqn
        \left. \sigma(\mchi) \right|_\sigma
 \=     \frac{\abrac{\mX / \mY}^{5 / 2} \vbrac{\mX - \mY}
              \abrac{\calR_{\sigma, X} / \calR_{\sigma, Y}} }
             {\bbrac{\calR_{\sigma, X} / \calR_{\sigma, Y} - \abrac{\mX / \mY}^{5 / 2}}^2}
        \non\\
 \conti ~~~~~~ \times %6
        \bbrac{  \frac{1}{\calR_{\sigma, X}^2}
                 \sum_{i, j = 2}^3
                 \aPp{\calR_{\sigma, X}}{c_{i, X}} \aPp{\calR_{\sigma, X}}{c_{j, X}}
                 {\rm cov}(c_{i, X}, c_{j, X})
               + (X \lto Y)}^{1 / 2}
\~,
\label{eqn:sigma_mchi_Rsigma}
\eeqn
 where I have used again
 the short--hand notation in Eq.~(\ref{eqn:ciX});
 note that $c_{1, (X, Y)} = I_{n, (X, Y)}$ do not appear here.
 Expressions for the derivatives of $\calR_{\sigma, X}$
 can be computed from Eq.~(\ref{eqn:RsigmaX_min}) as
\cheqnXa{A}
\beq
   \Pp{\calR_{\sigma, X}}{\IzX}
 = \bfrac{\FQminX}{2 \QminX^{1 / 2} r_X(\QminX) + \IzX \FQminX}
   \calR_{\sigma, X}
\~,
\label{eqn:dRsigmaX_dIzX}
\eeq
\cheqnXb{A}
\beq
   \Pp{\calR_{\sigma, X}}{r_X(\QminX)}
 = \bfrac{2 \QminX^{1 / 2}}{2 \QminX^{1 / 2} r_X(\QminX) + \IzX \FQminX}
   \calR_{\sigma, X}
\~;
\label{eqn:dRsigmaX_drminX}
\eeq
\cheqnX{A}%
 and similarly for the derivatives of $\calR_{\sigma, Y}$.
 Remind that
 factors $\calR_{\sigma, (X, Y)}$ appearing here
 can also be cancelled by the prefactors
 in the bracket in Eq.~(\ref{eqn:sigma_mchi_Rsigma}).

 In order to yield the best--fit WIMP mass
 as well as to minimize its statistical uncertainty
 by combining the estimators for different $n$
 in Eq.~(\ref{eqn:mchi_Rn}) with each other
 and with the estimator in Eq.~(\ref{eqn:mchi_Rsigma}),
 a $\chi^2$ function has been introduced as
 \cite{DMDDmchi}
\beq
   \chi^2(\mchi)
 = \sum_{i, j}
   \abrac{f_{i, X} - f_{i, Y}} {\cal C}^{-1}_{ij} \abrac{f_{j, X} - f_{j, Y}}
\~,
\label{eqn:chi2}
\eeq
 where
\cheqnXa{A}
\beqn
           f_{i, X}
 \eqnequiv \alpha_X^i
           \bfrac{  2 Q_{{\rm min}, X}^{(i+1)/2} r_X(\Qmin) / F^2_X(Q_{{\rm min}, X})
                  + (i+1) I_{i, X}}
                 {  2 Q_{{\rm min}, X}^{   1 /2} r_X(\Qmin) / F^2_X(Q_{{\rm min}, X})
                  +       \IzX}
           \afrac{1}{300~{\rm km/s}}^i
           \non\\
           \non\\
 \=        \afrac{\alpha_X {\cal R}_{i, X}}{300~{\rm km/s}}^{i}
\~,
\label{eqn:fiXa}
\eeqn
 for $i = -1,~1,~2,~\dots,~n_{\rm max}$, and
\cheqnXb{A}
\beqn
           f_{n_{\rm max}+1, X}
 \eqnequiv \calE_X
           \bfrac{A_X^2}
                 {  2 Q_{{\rm min}, X}^{1/2} r_X(\Qmin) / F^2_X(Q_{{\rm min}, X})
                  + \IzX}
           \afrac{\sqrt{\mX}}{\mchi + \mX}
           \non\\
           \non\\
 \=        \frac{A_X^2}{\calR_{\sigma, X}} \afrac{\sqrt{\mX}}{\mchi + \mX}
\~;
\label{eqn:fiXb}
\eeqn
\cheqnX{A}%
 the other $n_{\rm max} + 2$ functions $f_{i, Y}$
 can be defined analogously.
 Here $n_{\rm max}$ determines the highest moment of $f_1(v)$
 that is included in the fit.
 The $f_i$ are normalized such that
 they are dimensionless and very roughly of order unity
 in order to alleviate numerical problems
 associated with the inversion of their covariance matrix.
 Note that
 the first $n_{\rm max} + 1$ fit functions
 depend on $\mchi$ only through the overall factor $\alpha$
 and $\mchi$ in Eqs.~(\ref{eqn:fiXa}) and (\ref{eqn:fiXb})
 is now a fit parameter,
 which may differ from the true value of the WIMP mass.
 Finally,
 $\cal C$ in Eq.~(\ref{eqn:chi2}) is the total covariance matrix.
 Since the $X$ and $Y$ quantities
 are statistically completely independent,
 $\cal C$ can be written as a sum of two terms:
\beq
   {\cal C}_{ij}
 = {\rm cov}\abrac{f_{i, X}, f_{j, X}} + {\rm cov}\abrac{f_{i, Y}, f_{j, Y}}
\~.
\label{eqn:Cij}
\eeq
 The entries of the $\cal C$ matrix given here
 involving basically only
 the moments of the WIMP velocity distribution
 can be read off Eq.~(82) of Ref.~\cite{DMDDf1v},
 with an slight modification
 due to the normalization factor in Eq.~(\ref{eqn:fiXa})%
\footnote{
 Since the last $f_i$ defined in Eq.~(\ref{eqn:fiXb})
 can be computed from the same basic quantities,
 i.e., the counting rates at $\Qmin$ and the integrals $I_0$,
 it can directly be included in the covariance matrix.
}:
\beqn
        {\rm cov}\abrac{f_i, f_j}
 \=     \calN_{\rm m}^2
        \bbiggl{  f_i \~ f_j \~ {\rm cov}(I_0, I_0)
                + \Td{\alpha}^{i+j} (i+1) (j+1) {\rm cov}(I_i, I_j)}
        \non\\
 \conti ~~~~~~~~ %8
                - \Td{\alpha}^j (j+1) f_i \~ {\rm cov}(I_0, I_j)
                - \Td{\alpha}^i (i+1) f_j \~ {\rm cov}(I_0, I_i)\bigg.
        \non\\
 \conti ~~~~~~~~~~~~ %12
                + D_i D_j \sigma^2(r(\Qmin))
                - \abrac{D_i f_j + D_j f_i} {\rm cov}(r(\Qmin), I_0)\Bigg.
        \non\\
 \conti ~~~~~~~~~~~~~~~~ %16
        \bbiggr{+ \Td{\alpha}^j (j+1) D_i \~ {\rm cov}(r(\Qmin), I_j)
                + \Td{\alpha}^i (i+1) D_j \~ {\rm cov}(r(\Qmin), I_i)}
\~.
        \non\\
\label{eqn:cov_fi}
\eeqn
 Here I used
\beq
        \calN_{\rm m}
 \equiv \frac{1}{2 \Qmin^{1 /2} r(\Qmin)/\FQmin + I_0}
\~,
\label{eqn:calNm}
\eeq
\beq
        \Td{\alpha}
 \equiv \frac{\alpha}{300~{\rm km/s}}
\~,
\label{eqn:td_alpha}
\eeq
 and
\cheqnXa{A}
\beq
        D_i
 \equiv \frac{1}{\cal N_{\rm m}} \bPp{f_i}{r(\Qmin)}
 =      \frac{2}{\FQmin}
        \abigg{\Td{\alpha}^i \Qmin^{(i+1)/2} - \Qmin^{1/2} \~ f_i}
\~,
\label{eqn:Dia}
\eeq
 for $i = -1,~1,~2,~\dots,~n_{\rm max}$; and
\cheqnXb{A}
\beq
   D_{n_{\rm max}+1}
 = \frac{2}{\FQmin} \abrac{-\Qmin^{1/2} f_{n_{\rm max}+1}}
\~.
\label{eqn:Dib}
\eeq
\cheqnX{A}%

 Finally,
 since the basic requirement of the expressions for determining $\mchi$
 given in Eqs.~(\ref{eqn:mchi_Rn}) and (\ref{eqn:mchi_Rsigma}) is that,
 from two experiments with different target nuclei,
 the values of a given moment of the WIMP velocity distribution
 estimated by Eq.~(\ref{eqn:moments}) should agree,
% This means that
 the upper cuts on $f_1(v)$ in two data sets
 should be (approximately) equal%
\footnote{
 Here the threshold energies
 have been assumed to be negligible.
}.
 Since $v_{\rm cut} = \alpha \sqrt{Q_{\rm max}}$,
 it requires that \cite{DMDDmchi}
\beq
   Q_{{\rm max}, Y}
 = \afrac{\alpha_X}{\alpha_Y}^2 Q_{{\rm max}, X}
\~.
\label{eqn:match}  
\eeq
 Note that
 $\alpha$ defined in Eq.~(\ref{eqn:alpha})
 is a function of the true WIMP mass.
 Thus this relation for matching optimal cut--off energies
 can be used only if $\mchi$ is already known.
 One possibility to overcome this problem is
 to fix the cut--off energy of the experiment with the heavier target,
 minimize the $\chi^2(\mchi)$ function
 defined in Eq.~(\ref{eqn:chi2}),
 and then estimate the cut--off energy for the lighter nucleus
 by Eq.~(\ref{eqn:match}) algorithmically \cite{DMDDmchi}.
\subsection{Statistical uncertainty on \boldmath$|f_{\rm p}|^2$}
 By using the standard Gaussian error propagation,
 the statistical uncertainty on $|f_{\rm p}|^2$
 estimated by Eq.~(\ref{eqn:fp2}) can be given as
\beq
    \sigma(|f_{\rm p}|^2)
 = |f_{\rm p}|^2
    \bbrac{  \frac{\sigma^2(\mchi)}{(\mchi + \mN)^2}
           + \calN_{\rm m}^2 \sigma^2(1 / \calN_{\rm m})
           + \frac{2 \calN_{\rm m} \~ {\rm cov}(\mchi, 1 / \calN_{\rm m})}
                  {(\mchi + \mN)}}^{1/2}
\~.
\label{eqn:sigma_fp2}
\eeq
 Here the statistical error on $1 / \calN_{\rm m}$
 can be given from Eq.~(\ref{eqn:calNm}) directly as
\beq
      \sigma^2(1 / \calN_{\rm m})
 =    \bfrac{2 \Qmin^{1 / 2}}{\FQmin}^2 \sigma^2(r(\Qmin))
  +   \sigma^2(I_0)
  + 2 \bfrac{2 \Qmin^{1 / 2}}{\FQmin} {\rm cov}(r(\Qmin), I_0)
\~.
\label{eqn:sigma2_calNm}
\eeq
 For the case that
 one has only two data sets with different target nuclei, $X$ and $Y$,
 one of these two data sets will then be needed
 for reconstructing the WIMP mass $\mchi$ and
 also for estimating $1 / \calN_{\rm m}$ in Eq.~(\ref{eqn:fp2}).
 The uncertainties on $\mchi$ and $1 / \calN_{\rm m}$
 are thus correlated.
 Assuming that
 the WIMP mass is reconstructed by Eq.~(\ref{eqn:mchi_Rn}),
 and target $X (Y)$ is used for estimating $1 / \calN_{\rm m}$,
 the covariance of $\left. \mchi \right|_{\Expv{v^n}}$
 and $1 / \calN_{{\rm m}, (X, Y)}$ can be obtained
 by modifying Eq.~(\ref{eqn:sigma_mchi_Rn}) slightly as
\cheqnXa{A}
\beqn
 \conti {\rm cov}(\left. \mchi \right|_{\Expv{v^n}}, 1 / \calN_{{\rm m}, X})
        \non\\
 \=     \frac{\sqrt{\mX / \mY} \abrac{\mX - \mY} \abrac{\calR_{n, X} / \calR_{n, Y}} }
             {\abrac{\calR_{n, X} / \calR_{n, Y} - \sqrt{\mX / \mY}}^2}
        \afrac{1}{\calR_{n, X}}
        \non\\
 \conti ~~~~~~ \times %6
               \sum_{i = 1}^3
               \aPp{\calR_{n, X}}{c_{i, X}}
               \bbrac{  {\rm cov}(c_{i, X}, \IzX)
                      + {\rm cov}(c_{i, X}, r_X(\QminX))
                        \afrac{2 \QminX^{1 / 2}}{\FQminX} } %}
\~,
\eeqn
 and
\cheqnXb{A}
\beqn
 \conti {\rm cov}(\left. \mchi \right|_{\Expv{v^n}}, 1 / \calN_{{\rm m}, Y})
        \non\\
 \=     \frac{\sqrt{\mX / \mY} \abrac{\mX - \mY} \abrac{\calR_{n, X} / \calR_{n, Y}} }
             {\abrac{\calR_{n, X} / \calR_{n, Y} - \sqrt{\mX / \mY}}^2}
        \afrac{-1}{\calR_{n, Y}}
        \non\\
 \conti ~~~~~~ \times %6
               \sum_{i = 1}^3
               \aPp{\calR_{n, Y}}{c_{i, Y}}
               \bbrac{  {\rm cov}(c_{i, Y}, \IzY)
                      + {\rm cov}(c_{i, Y}, r_Y(\QminY))
                        \afrac{2 \QminY^{1 / 2}}{\FQminY} } %}
\~.
\eeqn
\cheqnX{A}%
 For the case that
 the WIMP mass is reconstructed by Eq.~(\ref{eqn:mchi_Rsigma}),
 one can also modify Eq.~(\ref{eqn:sigma_mchi_Rsigma})
 to obtain that
\cheqnXa{A}
\beqn
 \conti {\rm cov}(\left. \mchi \right|_\sigma, 1 / \calN_{{\rm m}, X})
        \non\\
 \=     \frac{\abrac{\mX / \mY}^{5 / 2} \abrac{\mX - \mY}
              \abrac{\calR_{\sigma, X} / \calR_{\sigma, Y}} }
             {\bbrac{\calR_{\sigma, X} / \calR_{\sigma, Y} - \abrac{\mX / \mY}^{5 / 2}}^2}
        \afrac{1}{\calR_{\sigma, X}}
        \non\\
 \conti ~~~~~~ \times %6
               \sum_{i = 2}^3
               \aPp{\calR_{\sigma, X}}{c_{i, X}}
               \bbrac{  {\rm cov}(c_{i, X}, \IzX)
                      + {\rm cov}(c_{i, X}, r_X(\QminX))
                        \afrac{2 \QminX^{1 / 2}}{\FQminX} } %}
\~,
\eeqn
 and
\cheqnXb{A}
\beqn
 \conti {\rm cov}(\left. \mchi \right|_\sigma, 1 / \calN_{{\rm m}, Y})
        \non\\
 \=     \frac{\abrac{\mX / \mY}^{5 / 2} \abrac{\mX - \mY}
              \abrac{\calR_{\sigma, X} / \calR_{\sigma, Y}} }
             {\bbrac{\calR_{\sigma, X} / \calR_{\sigma, Y} - \abrac{\mX / \mY}^{5 / 2}}^2}
        \afrac{-1}{\calR_{\sigma, Y}}
        \non\\
 \conti ~~~~~~ \times %6
               \sum_{i = 2}^3
               \aPp{\calR_{\sigma, Y}}{c_{i, Y}}
               \bbrac{  {\rm cov}(c_{i, Y}, \IzY)
                      + {\rm cov}(c_{i, Y}, r_Y(\QminY))
                        \afrac{2 \QminY^{1 / 2}}{\FQminY} } %}
\~.
\eeqn
\cheqnX{A}%
 Note that,
 firstly,
 in the above expressions
 we have to use $\abrac{\mX - \mY}$
 instead of $\vbrac{\mX - \mY}$
 in Eqs.~(\ref{eqn:sigma_mchi_Rn}) and (\ref{eqn:sigma_mchi_Rsigma});
 for expressions with the $Y$ target,
 there is an additional ``$-$ (minus)'' sign.
 Secondly,
 the algorithmic process for matching
 the experimental maximal cut--off energies of two experiments
 used for the reconstruction of the WIMP mass
 can also be used with the basic expressions
 (\ref{eqn:mchi_Rn}) and (\ref{eqn:mchi_Rsigma}).
 For this case and the lighter nucleus
 is used for estimating $1 / \calN_{\rm m}$,
 the energy range
 of the sum in Eq.~(\ref{eqn:cov_In}) or
 of the integral in Eq.~(\ref{eqn:cov_In_int})
 as the estimator for the covariance of $I_n$
 should be modified to be between $\Qmin$ and
 the {\em reduced} maximal cut--off energy
 of the lighter nucleus.
%
%
% Appendix B
\section{Proportionality of \boldmath$r_{\rm bg}(\Qmin)$ and $I_{0, \rm bg}$
         to $R_{\rm sg}$}
 The spectrum of residue background events
 {\em before} normalized by the experimental exposure
 $\calE = \calE_{\rm sg}$
 can be expressed as
\beq
   \adRdQ_{\rm bg, expt}
 = a \calE_{\rm sg} \adRdQ_{\rm bg}
\~.
\label{eqn:dRdQ_bg_expt}
\eeq
 Here $a$ is a proportional constant,  
 $\calE_{\rm sg}$ is
 the required exposure to observe
 the expected ``WIMP signal'' (not total) events,
 which can be estimated theoretically by
\beq
   \calE_{\rm sg}
 = \frac{N_{\rm sg}}{R_{\rm sg}}
 = \bBig{N_{\rm tot} (1 - r_{\rm bg})}
   \bbrac{\int_{\Qmin}^{\Qmax} \adRdQ_{\rm sg} dQ}^{-1}
\~,
\label{eqn:calE_sg}
\eeq
 where $N_{\rm tot}$ and $N_{\rm sg}$
 are the number of the total and WIMP signal events,
 respectively;
 $0 \le r_{\rm bg} \le 1$ is
 the ratio of residue background events
 in the whole data set.
 On the other hand,
 the number of residue background events
 in the data set can be given by
\beq
   N_{\rm bg}
 = N_{\rm tot} r_{\rm bg}
 = \int_{\Qmin}^{\Qmax}
   \adRdQ_{\rm bg, expt} dQ
 = a \calE_{\rm sg}
 = \int_{\Qmin}^{\Qmax}
   \adRdQ_{\rm bg} dQ
\~.
\label{eqn:N_bg}
\eeq
 $(dR / dQ)_{\rm bg}$ in Eq.~(\ref{eqn:dRdQ_bg_expt}) and here
 is a (simplified) analytic form of the background spectrum,
 e.g., $(dR / dQ)_{\rm bg, ex}$ in Eq.~(\ref{eqn:dRdQ_bg_ex})
 and $(dR / dQ)_{\rm bg, const}$ in Eq.~(\ref{eqn:dRdQ_bg_const}).

 Similar to Eq.~(\ref{eqn:In_int}),
 $I_{n, \rm bg}$ can be estimated from the background spectrum
 $(dR / dQ)_{\rm bg, expt}$ by
\beqn
    I_{n, \rm bg}(\Qmin, \Qmax)
 \= \int_{\Qmin}^{\Qmax}
    \frac{Q^{(n-1)/2}}{F^2(Q)} \adRdQ_{\rm bg, expt} dQ
    \non\\
 \= a \calE_{\rm sg}
    \int_{\Qmin}^{\Qmax}
    \frac{Q^{(n-1)/2}}{F^2(Q)} \adRdQ_{\rm bg} dQ
\~.
\label{eqn:I0_bg}
\eeqn
 Hence,
 the second term involving
 $r_{\rm bg}(\Qmin)$ and $I_{0, \rm bg}(\Qmin, \Qmax)$
 on the right--hand side of Eq.~(\ref{eqn:rmin_I0})
 can be given as
\beqn
 \conti     \frac{1}{\calE_{\rm sg}}
            \bbrac{\frac{2 \Qmin^{1/2} r_{\rm bg}(\Qmin)}{\FQmin} + I_{0, \rm bg}}
            \non\\
 \=         a
            \bbrac{  \frac{2 \Qmin^{1/2}}{\FQmin} \adRdQ_{{\rm bg}, \~ Q = \Qmin}
                   + \int_{\Qmin}^{\Qmax}
                     \frac{1}{\sqrt{Q} \~ F^2(Q)} \adRdQ_{\rm bg} dQ}
\~.
%\label{eqn:}
\eeqn
 Finally,
 by combining Eqs.~(\ref{eqn:calE_sg}) and (\ref{eqn:N_bg}),
 the proportional constant $a$ can be calculated by
\beqn
            a
 \=         \frac{N_{\rm tot} r_{\rm bg}}{\calE_{\rm sg}}
            \bbrac{\int_{Q_{\rm min, bg}}^{Q_{\rm max, bg}} \adRdQ_{\rm bg} dQ}^{-1}
            \non\\
 \=         \frac{r_{\rm bg}}{1 - r_{\rm bg}}
            \bbrac{\int_{\Qmin}^{\Qmax} \adRdQ_{\rm sg} dQ}
            \bbrac{\int_{Q_{\rm min, bg}}^{Q_{\rm max, bg}} \adRdQ_{\rm bg} dQ}^{-1}
            \non\\
 \eqnpropto \afrac{r_{\rm bg}}{1 - r_{\rm bg}}
            R_{\rm sg}(\Qmin, \Qmax)
\~.
\label{eqn:coe_a}
\eeqn
 Remind that,
 while the signal spectrum $(dR / dQ)_{\rm sg}$
 is a function of the WIMP mass $\mchi$,
 the background spectrum $(dR / dQ)_{\rm bg}$
 should in general be {\em independent} of $\mchi$.
\end{document}